\documentclass[aps,twocolumn,showpacs,preprintnumbers,floatfix,nofootinbib]{revtex4}
\usepackage{graphicx}
\usepackage{graphics}
\usepackage[subfigure]{graphfig}
\usepackage{epsfig}
\usepackage{epstopdf}
\usepackage{dcolumn}
\usepackage{amsmath}
\usepackage{color}

\def\be{\begin{equation}}
\def\ee{\end{equation}}
\def\bea{\begin{eqnarray}}
\def\eea{\end{eqnarray}}
\newcommand{\Tr}{\mathrm {Tr}}
\newcommand{\qe}{\mathrm {\!\!\!\!}}
\newcommand{\qqe}{\mathrm {\qe\qe\qe\qe}}

\definecolor{green}{rgb}{0,.5,0}

\begin{document}

\title{\vspace{1.0in} {\bf  Stochastic method with low mode
substitution for nucleon isovector matrix elements}}


\author{Yi-Bo Yang$^{1}$, Andrei Alexandru$^{2}$,  Terrence Draper$^{1}$,  Ming Gong$^{3}$, and
Keh-Fei Liu$^{1}$
\vspace*{-0.5cm}
\begin{center}
\large{
\vspace*{0.4cm}
\includegraphics[scale=0.20]{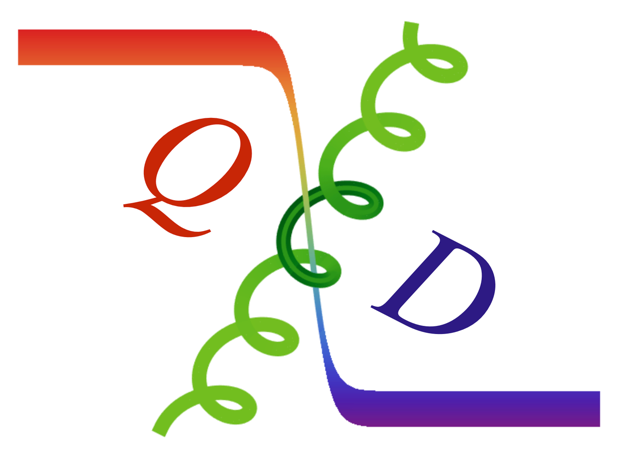}\\
\vspace*{0.4cm}
($\chi$QCD Collaboration)
\vspace*{0.4cm}
}
\end{center}
}
\affiliation{
$^{1}$Department of Physics and Astronomy, University of Kentucky, Lexington, KY 40506, USA\\
$^{2}$Department of Physics, The George Washington University, Washington, DC 20052, USA\\
$^{3}$Institute of High Energy Physics and Theoretical Physics Center for Science Facilities,\\
Chinese Academy of Sciences, Beijing 100049, China\\
}

\begin{abstract}
We introduce a stochastic method with low-mode substitution to evaluate the connected three-point functions. 
The isovector matrix elements of the nucleon for the axial-vector coupling $g_A^3$, scalar couplings $g_S^3$ and the quark momentum fraction $\langle x\rangle_{u -d}$ are calculated with overlap fermion on 2+1 flavor domain-wall configurations on a $24^3 \times 64$ lattice at $m_{\pi} = 330$ MeV with lattice spacing $a = 0.114$ fm. 
\end{abstract}

 \maketitle

\section{Introduction}

 
 The proton isovector-axial coupling $g_A^3$ and quark momentum fraction 
$\langle x \rangle_{u-d}$ are important benchmarks {to check whether the systematic uncertainties of lattice QCD simulation, such as finite lattice spacing, finite volume, and chiral extrapolation, are under control, by a correct reproduction of the corresponding experimental results.} Since the {noisy} disconnected  insertion contribution 
 to the isovector part of the nuclear matrix element is canceled between two degenerate flavors, the values are obtained solely from the connected insertion and thus are relatively {cheaper} to compute with high precision to be considered as benchmarks.

  Most attempts have resulted in values $\sim$10\% below the experimental number for the 
axial-vector coupling~\cite{Owen:2012ts,Bhattacharya:2013ehc,Alexandrou:2013joa,Alexandrou:2010hf,Ohta:2013qda,Bratt:2010jn,Syritsyn:2014xwa,Green:2012ud}, while a few claim that their results could be consistent with experiment~\cite{Capitani:2012gj,Horsley:2013ayv,Bali:2014nma,Abdel-Rehim:2015owa}. 
 For the quark momentum fraction $\langle x \rangle_{u-d}$, overestimation by $\sim$20 -- 30\% is common in most of the calculations \cite{Alexandrou:2013joa,Aoki:2010xg,Bali:2014gha,Pleiter:2011gw,Syritsyn:2014xwa} except \cite{Green:2012ud}.

Recently, attention has been paid to lattice QCD calculation of the isovector scalar matrix element $g_S^3$ in the proton \cite{Bhattacharya:2013ehc,Bali:2014nma,Green:2012ej,Gonzalez-Alonso:2013ura} due to its role in constraining possible scalar interactions at the TeV scale \cite{Bhattacharya:2011qm}.

In this work, we calculate the isovector matrix elements of the nucleon for the axial-vector and scalar couplings  and the quark momentum fraction with the valence overlap fermion on $2 +1$ flavor domain-wall fermion (DWF) configurations \cite{Aoki:2010dy}. Compared to simulations with other actions, the overlap fermion  provides the best control of the systematic errors {since it is free of} {explicit} chiral symmetry {breaking} and gives small $O(a^2)$ errors, whereas the  numerical work is more costly. 

In order to improve SNR, the  8-grid smeared $Z_3$ noise source with low-mode substitution (LMS) {\cite{DeGrand:2004qw,Giusti:2004yp,Giusti:2006mh,Foley:2005ac,Kaneko:2007nf}} has been applied to the hadron two point correlator on the $24^3 \times 64$ 
lattice~\cite{Gong:2013vja} which improves the error of the nucleon mass of a point source by a factor of 7 and
that of the 8-grid source without smearing by a factor of 2.5. In this work, we use a stochastic sandwich contraction method to remove the need of multiple inversions
in the sink-sequential approach and use the current-sequential method for the low modes in the propagator
between the current and the sink. This is an extension of the noise grid smeared source with LMS to the three point function. Such a many-to-all correlator with LMS  is useful when the low-eigenmode contributions are important in the relevant time windows where the physical quantities are extracted.

The structure of the rest of the paper is organized as follows. The LMS technique with noise grid source for the non-zero momentum case of the two point correlation function is provided in Sec.~\ref{sec:lms}.  Sec.~\ref{sec:ss} discusses the possibility of applying LMS on all the four quark propagators in the proton three-point function.  The numerical details are provided in Sec.~\ref{sec:numerical}. In Sec.~\ref{sec:res}, the results of  isovector matrix 
elements of the nucleon for the axial-vector $g_A^3$, the scalar coupling $g_S^3$ and the quark momentum fraction $\langle x \rangle_{u-d}$ are provided. A short summary and outlook are presented in Sec.~\ref{sec:summary}.

\section{Low mode substitution with mixed momentum grid source}\label{sec:lms}

Let's first consider the nucleon {two-point function (2pt) with the interpolation field of the nucleon \cite{Wilcox:1991cq},
\bea
\chi_{\alpha} (x) &=& \epsilon^{abc}\psi_\alpha^{(u) a}(x)\psi_\beta^{(u) b} (x) (\tilde{C})_{\beta\gamma}\psi_\gamma^{(d) c}(x) \nonumber\\
 \overline{\chi}_{\alpha'}(x)&=& -\epsilon^{a'b'c'}\overline{\psi}_ {\gamma'}^{(d) c'} (\tilde{C})_{\gamma' \beta'} \overline{\psi}_{\beta'}^{(u) b'}(x) \overline{\psi}_{\alpha'}^{(u) a'}(x),
\eea
 where $\tilde{C} \equiv C\gamma_5= \gamma_2 \gamma_4 \gamma_5$ in the Pauli-Sakurai gamma-matrix convention, used throughout this work. There are two kinds of the Wick contractions 
 so the 2pt of the nucleon can be constructed in terms of the point-to-point quark propagator $S$ as
\bea
&&\!\!\!\!\!C(y, x; \Gamma; S^{(u)}, S^{(d)}, S^{(u)}) = \langle \epsilon^{abc} \epsilon^{a^\prime b^\prime c^\prime} \nonumber\\
&&\Tr\left( \Gamma S^{(u)aa^\prime}(y, x) \right) \Tr\left( \underline{S}^{(d)bb^\prime}(y, x) S^{(u)cc^\prime}(y, x)\right) \rangle\nonumber\\
&&- \langle \epsilon^{abc} \epsilon^{a^\prime b^\prime c^\prime} \Tr\left( \Gamma S^{(u)ab^\prime}(y, x) \underline{S}^{(d)ba^\prime}(y, x) S^{(u)cc^\prime}(y, x)\right)\rangle\nonumber\\
&&\!\!\!\!\!=\langle \epsilon^{abc} \epsilon^{a^\prime b^\prime c^\prime} \nonumber\\
&&\Tr\left( \Gamma S^{(u)aa^\prime}(y, x) \right) \Tr\left( \underline{S}^{(d)bb^\prime}(y, x) S^{(u)cc^\prime}(y, x)\right)\nonumber\\
&&+\Tr\left( \Gamma S^{(u)aa^\prime}(y, x) \underline{S}^{(d)bb^\prime}(y, x) S^{(u)cc^\prime}(y, x)\right)\rangle
\label{eq:nuc00}
\eea
where $\underline{S}$ is defined as $(\tilde{C}S\tilde{C}^{-1})^T$} and
$\Gamma$ is the projection operator for the nucleon polarization.

The quark propagator $S$ in the above equation is the inverse of the operator $(D_c + m)$~\cite{Chiu:1998eu,Liu:2002qu}, where  $D_c$ is defined in terms of the overlap operator and is chiral, i.e. $\{D_c, \gamma_5\} = 0$ \cite{Chiu:1998gp}. The details will be discussed in Sec.~\ref{sec:numerical}. As in Ref.~\cite{Li:2010pw,Gong:2013vja}, we use the low lying eigenvalues and eigenvectors of the overlap fermion, $\lambda_i$ and $|i\rangle$, satisfying  $D_{c}|i\rangle=\lambda_i|i\rangle$ to speed up the inversion
and separate the propagator into its low-mode and high-mode parts,
\bea
S_L(y,x)&=&\sum_{|\lambda_i|<\epsilon_{c}}\frac{1}{\lambda_i +m}| i\rangle_y \langle i|_x,\nonumber\\
S_H(y,x)&=&S(y,x)-\sum_{|\lambda_i|<\epsilon_{c}}\frac{1}{\lambda_i + m}|i\rangle_y \langle i|_x,
\eea
with $\epsilon_{c}$ as the upper bound of the modulus of the eigenvalues.

The idea of using the $Z_3$ noise grid source is to tie the sources of the three quark
 propagators stochastically to each point (or a smeared point) on the grid so that one can have a multi-to-all 
 correlator from one inversion.
LMS for the quark propagator with $Z_3$  noise grid source (PropNG), be it point-grid  (PG)~\cite{Li:2010pw}
 or smeared grid  (SG)~\cite{Gong:2013vja}, has been used to improve the SNR for the nucleon correlator with significant success. This technique removes the gauge non-invariant contributions of the low-mode contributions (defined below) from the cases in which three propagators are from different source sites, and restores the benefit of using PropNG.

To construct the nucleon correlation function with LMS, PropNG $S_{NG}(y)$ should be split into its high-mode and low-mode pieces
\bea
 S_{NG}(y)&=&\sum_{x\in G} \theta(x) S(y,x)\nonumber\\
&=&S^H_{NG}(y)+\sum_{x\in G}\theta(x)S^{L}(y,x)\label{eq:high_low},
\eea
with $S^H_{NG}(y)=\sum_{x\in G}\theta(x)S^{H}(y,x)$ and random $Z_3$ phases $\theta(x)\in \{1,e^{i\frac{2}{3}\pi},e^{-i\frac{2}{3}\pi}\}$ for each point 
 on a grid $G$. 

As in Ref.~\cite{Gong:2013vja}, we can expand the nucleon correlation function 
$C(y, x; \Gamma; S_{NG}^{(u)}, S_{NG}^{(d)}, S_{NG}^{(u)})$ with the decomposition in Eq.~(\ref{eq:high_low}) (ignoring the indices for the sink position $y$ and the projection matrix $\Gamma$), 

\begin{widetext}
\bea  \label{eq:lms_2pt_all}
&&C^{LMS}\big(S_{NG},S_{NG},S_{NG}\big) 
= 
\nonumber \\
&=& { C(S^H_{NG}, S^H_{NG}, S^H_{NG})} + {\sum_{x\in G} C\big(\theta(x) S^L(x), \theta(x) S^L(x), \theta(x) S^L(x)\big) }\nonumber \\
           &&+ {C\big(\sum_{x\in G} \theta(x)S^L(x), S^H_{NG}, S^H_{NG}\big) + C\big(S^H_{NG}, \sum_{x\in G} \theta(x)S^L(x), S^H_{NG}\big) + C\big(S^H_{NG}, S^H_{NG}, \sum_{x\in G} \theta(x)S^L(x)\big)} \nonumber \\
           &&+ {\sum_{x\in G} C\big(\theta(x) S^L(x), \theta(x) S^L(x), S^H_{NG}\big) + \sum_{x\in G} C\big(\theta(x) S^L(x), S^H_{NG}, \theta(x) S^L(x)\big) + \sum_{x\in G} C\big(S^H_{NG}, \theta(x) S^L(x), \theta(x)S^L(x)\big)} \nonumber \\
 &=& {C_{ker}\big(S^H_{NG},\sum_{x\in G} \theta(x)S^L(x)\big) }+ {\sum_{x\in G}  C_{ker}\big(\theta(x)S^L(x), S^H_{NG}\big)} 
\eea
\end{widetext} 
where 
\bea   \label{eq:lms_2pt_ker}
C_{ker}(S_1,S_2)&=&  C(S_1,S_1,S_1)+ C(S_2,S_1,S_1)  \nonumber\\
&& + C(S_1,S_2,S_1) + C(S_1,S_1,S_2).
\eea

The nucleon correlator with LMS here can be obtained from the one in Ref.~\cite{Gong:2013vja} with just one more step. The low-mode propagator ${ \sum}_{x\in G}\theta(x)S^{L}(y,x)$ is decomposed into several terms as in the very last term in the RHS of Eq.~\ref{eq:lms_2pt_all} to improve the SNR. 

After the noise averaging, the nucleon correlation function with PropNG should be
a stochastic estimate of the sum of nucleon correlators from each of the grid points, i.e.

\bea
\sum_{\vec{y}}C_{grid}({\vec{y}})&=&\sum_i\sum_{\vec{y}}C({\vec{y}},{\vec{w}}_i), 
\eea
where the grid points $\vec{w}_i$ are
\be
\vec{w}_i \in (x_0+m_x\Delta_x ,y_0+m_y\Delta_y,z_0+m_z\Delta_z).
\ee 
with $m_{x,y,z} = (0, 1,\cdots, L_s/\Delta_{x,y,z})$ modulo the periodic boundary 
condition in the spatial directions. In this grid pattern, in addition to the zero momentum mode (0,0,0), one can obtain  non-zero momentum modes from the nucleon correlation function with PropNG. For example, for the  PropNG with a regular ($\Delta_x=\Delta_y=\Delta_z=L_s/m$) grid, the momentum mode $p=(\pm{n_1m}, \pm{n_2m}, \pm{n_3m})$ ($n_{1,2,3}$ are integers) can be obtained. In this case, there is a phase factor which needs to be taken into account when the origin 
$w_0=(x_0, y_0, z_0)$ is changed from configuration to configuration,
 \bea
 \sum_{\vec{y}} &&\!\!\!\!\!\!C_{grid}({\vec{y}}) e^{- i\frac{2 \pi}{L_s}\vec{y}\cdot p} 
=e^{-i \frac{2n \pi}{L_s}w_0\cdot p}\nonumber\\
&&\sum_i\sum_{\vec{y}}C({\vec{y}},{\vec{w}}_i)e^{- i\frac{2 \pi}{L_s}(\vec{y}-\vec{w_i})\cdot p- i\frac{2 m\pi}{L_s}(\vec{w_i}-\vec{w_0})\cdot(n_1 ,n_2,n_3)}\nonumber\\
&=&e^{-i \frac{2n \pi}{L_s}w_0\cdot p}\sum_i\sum_{\vec{y}}C({\vec{y}},{\vec{w}}_i)e^{- i\frac{2 \pi}{L_s}(\vec{y}-\vec{w_i})\cdot p}
 \eea
 The exponential term in the second line with the exponent proportional to $\vec{w_i}-\vec{w_0}$ does not contribute, since all components of the latter are proportional to $L_s/m$ and, as a result, the exponent is a multiple of $2\pi$.
  
In order to obtain the other momentum modes, propagators with noise grid non-zero momentum source (PropNGM) are required. To cover a range of $p^2$ modes and minimize the effect of the rotation symmetry breaking due to the finite lattice spacing and volume, three kinds of PropNGM
 \bea
 &&S_{p_1}(y)=\sum_i\theta({\vec{{w}}_i})S({\vec{y}},{\vec{{w}}}_i)e^{i\frac{2 \pi}{L_s}\vec{{w}_i}\cdot(1,0,0)},\nonumber\\
 &&S_{p_2}(y)=\sum_i\theta({\vec{{w}}_i})S({\vec{y}},{\vec{{w}}}_i)e^{i\frac{2 \pi}{L_s}\vec{{w}_i}\cdot(0,1,0)},\nonumber\\
&&S_{p_3}(y)=\sum_i\theta({\vec{{w}}_i})S({\vec{y}},{\vec{{w}}}_i)e^{i\frac{2 \pi}{L_s}\vec{{w}_i}\cdot(0,0,1)}
 \eea
 and related inversions are required for the proton case. It is trivial to confirm that one can obtain a momentum mode like (1,1,0) from the contraction $C(S_{p_1},S_{p_2},S_{NG})$, and (1,1,1) from $C(S_{p_1},S_{p_2},S_{p_3})$. 

To reduce the cost, we can combine these three kinds of PropNGM together as the mixed PropNGM,
\bea
S_{p}&\equiv& S_{p_1}+S_{p_2}+S_{p_3}\nonumber\\
&=&\sum_i\theta({\vec{{w}}_i})S({\vec{y}},{\vec{{w}}}_i)(e^{i\frac{2 \pi}{L_s}\vec{{w}_i}\cdot(1,0,0)}\nonumber\\
&&+e^{i\frac{2 \pi}{L_s}\vec{{w}_i}\cdot(0,1,0)}+e^{i\frac{2 \pi}{L_s}\vec{{w}_i}\cdot(0,0,1)}),
\eea
with the origin of the grid $\vec{{w}}_0=(x_0,y_0,z_0)$ to be selected randomly for each configuration.

\begin{figure}[tbh]
\begin{center}
\includegraphics[scale=0.7]{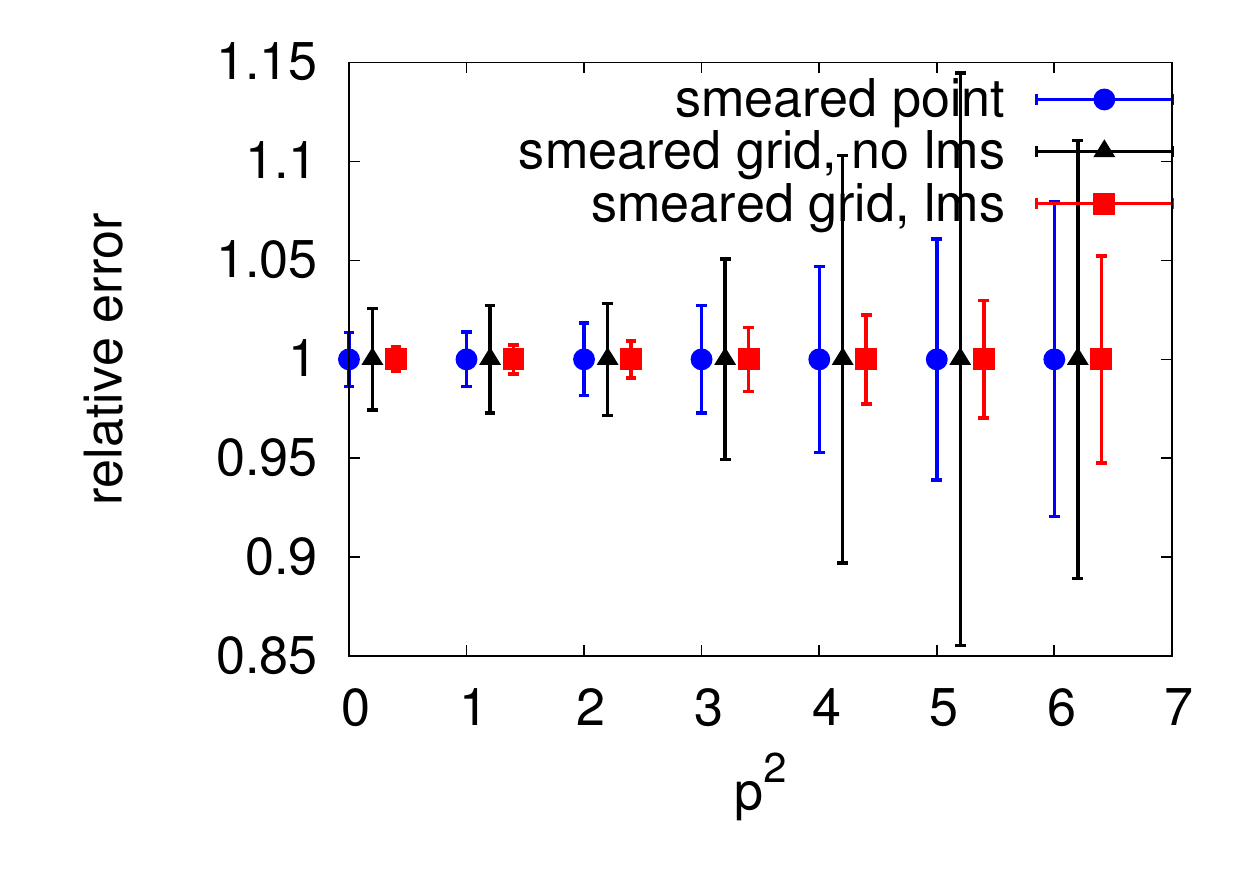}
\caption{\small The plot shows the relative error of 2pt as a function of the momentum squared $p^2$ at t=8  in lattice units. The data points of the smeared grid cases have been shifted a bit on the abscissa to make it easier to distinguish them. The SNR of the case with the noise smeared grid source (red squares) and LMS applied is better than the one with smeared point source (blue dots), while the one with the noise smeared grid source but no LMS (black triangles) is even worse than the one with smeared point source.\label{fig:2pt_error}}
\end{center}
\end{figure}

Fig.~\ref{fig:2pt_error} shows the SNR of the proton effective mass at the unitary point where the pion mass due to the valence quark is the same as that from the sea, {on the ensemble of which details will be addressed in Sec. \ref{sec:numerical}}. When LMS is applied, the SNR of the 2pt with the noise smeared grid source propagators (PropNG and mixed PropNGM, $\Delta_x=\Delta_y=\Delta_z=L_s/2$) is 
2.3 times smaller than that of the of the smeared point source at ${ p}^2=0$. This is a gain of 5.3 in statistics which is very good considering that the maximum possible gain is 8 for the ideal case where the independent nucleon propagators emerge from each of the 8 smeared grid points. On the other hand, if we don't use LMS, the SNR of 2pt with grid source is worse than the smeared point source, even though the latter has only 1/8 of the statistics of the former.
This is understood as due to the fact that the Parisi-Lepage estimate of the SNR for the nucleon is modified to
{
\begin{equation}  \label{s2n_N}
\frac{C_N(t,\vec{p}=0)}{\sigma_N(t)} \approx \sqrt{\frac{N}{V_3}} e^{-(m_N-3/2 m_{\pi})t},
\end{equation}
where  $N$ is the product of the number of noise and the number of gauge configurations and 
$V_3$ is the three-volume of the noise with its support on a time slice. In our case, $V_3 = 8$. It is this extra factor of
$\frac{1}{\sqrt{V_3}}$ which makes the SRN of the 2pt from the noise smeared grid source without LMS worse than
that of the smeared point source. When LMS is employed, the situation is reversed and one gains a statistical factor
almost as large as the number of the { grid} points. Thus, it is essential to have LMS when the noise grid source is
used for the nucleon.  

\bigskip
\section{LMS of the connected three-point correlator}\label{sec:ss}

Generally, a nucleon three point function (3pt), from $x$ to $y$, with a current {$\bar{\psi}(x)^{(u)}{\cal O}(z)\psi(x)^{(u)}$ (with current operator ${\cal O}$ such as $\gamma_i$, 
$\gamma_iD_j$, etc.) inserted at $z$, 
includes four kinds of Wick contractions,
\begin{widetext}
\bea
C_3^u(y,x;\Gamma; \hat{\cal S}^{(u)}, S^{(u)}, S^{(d)}, S^{(u)})&=&\langle \epsilon^{abc} \epsilon^{a^\prime b^\prime c^\prime} 
\Tr\left( \Gamma S^{(u)ad}(y, z){\cal O}(z)S^{(u)da^\prime}(z, x) \right) \Tr\left( \underline{S}^{(d)bb^\prime}(y, x) S^{(u)cc^\prime}(y, x)\right) \rangle\nonumber\\
&&+\langle \epsilon^{abc} \epsilon^{a^\prime b^\prime c^\prime} \Tr\left( \Gamma S^{(u)ad}(y, z){\cal O}(z)S^{(u)da^\prime}(z, x)  \underline{S}^{(d)bb^\prime}(y, x) S^{(u)cc^\prime}(y, x)\right)\rangle\nonumber\\
&&+\langle \epsilon^{abc} \epsilon^{a^\prime b^\prime c^\prime} 
\Tr\left( \Gamma S^{(u)aa^\prime}(y, x) \right) \Tr\left( \underline{S}^{(d)bb^\prime}(y, x) S^{(u)cd}(y, z){\cal O}(z)S^{(u)dc^\prime}(z, x)\right) \rangle\nonumber\\
&&+\langle \epsilon^{abc} \epsilon^{a^\prime b^\prime c^\prime} \Tr\left( \Gamma S^{(u)aa^\prime}(y, x) \underline{S}^{(d)bb^\prime}(y, x) S^{(u)cd}(y, z){\cal O}(z)S^{(u)dc^\prime}(z, x) \right)\rangle
\eea
\end{widetext}
and can be expressed in terms of the 2pt correlation function $C(y,x;\Gamma; S^{(u)}, S^{(d)}, S^{(u)})$ defined in Eq.~(\ref{eq:nuc00}),
\bea     \label{eq:3pt}
C_3^u&&\qe\qe\qe(y,x;\Gamma; \hat{\cal S}^{(u)}, S^{(u)}, S^{(d)}, S^{(u)}) \nonumber\\&=&C(y,x;\Gamma; \hat{\cal S}^{(u)}, S^{(d)}, S^{(u)})
\nonumber\\&+&C(y,x;\Gamma; S^{(u)}, S^{(d)}, \hat{\cal S}^{(u)}),
\eea
where $\hat{\cal S}({\cal O},z_0; y,x)\equiv \sum_{\vec{z}} S(y,z){\cal O}(z)S(z,x)$ is the current inserted propagator (PropCI). Similarly, the 3pt with a current of $d$ quark can be expressed as
\bea
C_3^d&&\qe\qe\qe(y,x;\Gamma; \hat{\cal S}^{(d)},  S^{(u)}, S^{(d)}, S^{(u)}) \nonumber\\&=&C(y,x;\Gamma; S^{(u)}, \hat{\cal S}^{(d)}, S^{(u)}).
\eea 
Fig.~\ref{fig:quark_diagram} shows PropCI as the product of the propagators in the shadowed region.}

\begin{figure}[tbh]
\begin{center}
\includegraphics[scale=0.2]{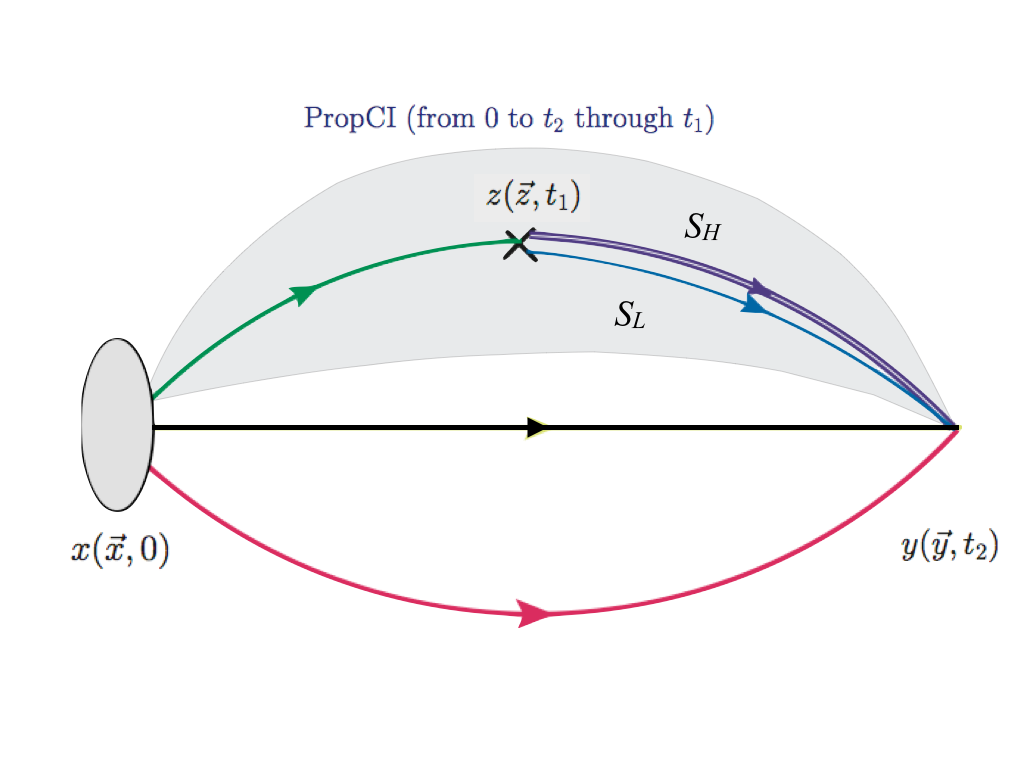}
{\caption{\small The quark diagram of the proton correlation function with the connected insertion, from $x$ to $y$, with an insertion at $z$. The product of the propagators in the shadowed region is the current inserted propagator, $\hat{\cal S}$. The propagator from the current $z$ to the sink $y$ is decomposed into its low- and high-mode contributions ($S_L$ and $S_H$ respectively) for further SNR/cost improvement from the advanced technique in the latter discussion. See Sec. \ref{sec:ssm} for more details.
\label{fig:quark_diagram}}}
\end{center}
\end{figure}

Supposing  $S^{(u)}=S^{(d)}=S$, Eq.~\ref{eq:3pt} can be rewritten into the contraction of PropCI $\hat{\cal S}$ and the remaining parts denoted as $X_{u,d}(\Gamma,S_1,S_2)$,

\bea
&&C_3^u(\Gamma;\hat{\cal S}, S,S,S) = \langle \Tr\big(\hat{\cal S} X_{u}(\Gamma,S,S)\big)\rangle, \nonumber\\
&&C_3^d(\Gamma;\hat{\cal S}, S,S,S) = \langle \Tr\big(\hat{\cal S} X_{d}(\Gamma,S,S)\big)\rangle,
\eea
with 
\bea\label{eq:X_etc}
X^{aa^\prime}_u(\Gamma,S_1,S_2)&=& \epsilon^{abc} \epsilon^{a^\prime b^\prime c^\prime} \nonumber\\
&&\big(
 \Gamma \textrm{Tr}[\underline{S_2}^{bb^\prime}S_1^{cc^\prime}] 
 +\underline{S_2}^{bb^\prime} S_1^{cc^\prime} \Gamma\nonumber\\
 &&\ \ \ \ 
+ \textrm{Tr}[\Gamma S_1^{cc^\prime}] \underline{S}_2^{bb^\prime} 
 +\Gamma S_1^{cc^\prime} \underline{S}_2^{bb^\prime}
 \big),\nonumber\\
X^{bb^\prime}_d(\Gamma,S_1, S_2)&=& \epsilon^{abc} \epsilon^{a^\prime b^\prime c^\prime} \nonumber\\
&&\big( \textrm{Tr}[\Gamma S_1^{aa^\prime}] \tilde{C}^{-1}({S_2}^{cc^\prime})^T\tilde{C}\nonumber\\
&&\ \ \ \  +\tilde{C}^{-1} ({S_1}^{aa^\prime}\Gamma S_2^{cc^\prime})^T \tilde{C} \big)
\eea

Based on the above definition, a typical 3pt correlation function for a point source on the $t=0$ time slice, when summed over the spatial indices of $y$ and $z$ becomes
\bea
C_3(t_2,t_1)&=&\sum_{\vec{y}}  
\big\langle\textrm{Tr}[ \hat{\cal S}({\cal O},t_1; \vec{y},t_2,\vec{0},0)\nonumber\\
&& X_{u,d}(\vec{y},t_2,\vec{0},0;\Gamma,S,S)] \big\rangle.
\eea

\subsection{Sink-sequential method and Stochastic sandwich method}

The typical problem of the connected 3pt is calculating the propagator from the current to the sink $S(\vec{y},t_2,\vec{z},t_1)$. On the surface, it is an all-to-all propagator which would be beyond the ability of the standard lattice inversion operation. 

However, when the sink time $t_2$ is fixed, the sequential source method \cite{Bernard:1985tm,Martinelli:1988rr} could be used, with $\gamma_5 X^{\dagger}_{u,d} (\vec{y},t_2,\vec{0},0) \gamma_5 $ as the source of the matrix inversion, to construct
\bea  \label{seq}
S_{seq}(X_{u,d};\vec{z},t_1,t_2,\vec{0},0)&=& \sum_{\vec{y}} S(\vec{z},t_1,\vec{y},t_2)\nonumber\\
&&\gamma_5  X^{\dagger}_{u,d} (\vec{y},t_2,\vec{0},0)\gamma_5.
\eea
Then, one can contract $S_{seq}$ with the standard quark propagator from $t=0$ to $t_1$ to construct the 3pt 
correlator,
\bea
C_3(t_2,t_1,{\cal O})&=&\sum_{\vec{z}, i}  \textrm{Tr}[\gamma_5S^{\dagger}_{seq}(X_{u,d},\vec{z},t_1,t_2,\vec{0},0)\gamma_5\nonumber\\
&&{\cal O}(\vec{z},t_1)S(\vec{z},t_1,\vec{0},0)],
\eea
taking the advantage of the relation ${ \gamma}_5 S(z,y)^\dagger \gamma_5 = S(y,z)$.

The disadvantage of the sequential method is that it has to calculate the sink-sequential propagator repeatedly when $X$ is changed for any reason, such as for: different  momentum, different quark flavor or mass, or different polarization projection of the baryon. This is expensive when many momenta are needed.

The number of inversions required in the sink-sequential method is $2 \times 4 \times N_p$ where
the 2 is for the $u$ and $d$ flavors in the nucleon, 4 is for the polarization, and $N_p$ is the number
of momentum projections. When many $N_p$ are required for nucleon form factors with momentum transfer (hundreds are needed for $|\vec{p}| \le 3$ with high statistics), the cost can be staggering.

A stochastic method~\cite{Evans:2010tg,Bali:2013gxx,Alexandrou:2013xon} (referred to as the stochastic sandwich method (SSM) in this work) is introduced to reduce the cost of the sequential method
when many sequential inversions are required. It entails inserting a noise estimate of the delta function 
$\delta(\vec{y_1}, \vec{y_2})$ at $t=t_2$,

\bea              \label{eq:3pt_sw}
\frac{1}{N_\text{noi}} \sum_{i=1}^{N_\text{noi}} &&\!\!\!\!\!\!\!\!\sum_{\vec{y}_1,\vec{y}_2,\vec{z}}  \textrm{Tr}\big{[}\theta^{(i)}_{\vec{y}_1}S(\vec{y}_1,t_2,\vec{z},t_1){\cal O}(\vec{z},t_1)S(\vec{z},t_1,\vec{0},0)\nonumber\\
&&X(\vec{0},0,\vec{y}_2,t_2)\theta^{(i)\dagger}_{\vec{y}_2}\big{]} {}_{\overrightarrow{N_{noi}\rightarrow\infty}}{C_3}(t_2,t_1,{\cal O}),
\eea
where {$N_{noi}$ is the number of the noises and }the noise $\theta$ satisfies
\be
\frac{1}{N_\text{noi}} \sum_{i=1}^{N_{noi}}\theta^{(i)}_{\vec{y}_1}\theta^{(i)\dagger}_{\vec{y}_2} {}_{\overrightarrow{N_{noi}\rightarrow\infty}} \delta_{\vec{y}_1,\vec{y}_2}. 
\ee
In other words, it uses the noise estimate of the all-to-all propagator, 
\bea
&&S(\vec{y}_1,t_2,\vec{z},t_1)\cong\sum_i\theta^{(i)}_{\vec{y}_1} \gamma_5(S^{{(i)}}_{noi}(\vec{z},t_1,t_2))^\dagger\gamma_5 
\eea
with
\bea
  S^{(i)}_{noi}(\vec{z},t_1,t_2)=\sum_{\vec{y}_1}S(\vec{z},t_1,\vec{y}_1,t_2)\theta^{(i)\dagger}, \label{eq:noi_prop}
\eea
instead of the original $S(\vec{y},t_2,\vec{z},t_1)$, to avoid the expensive calculation to construct the sink-sequential propagator with inversion of $2 \times 4 \times N_p$ sources. 

\subsection{Stochastic sandwich method (SSM) with LMS}\label{sec:ssm}
 
SSM avoids the cost of the repeated inversion for many different sequential sources, but it still requires multiple inversions for several noises, before the SNR can reach its upper limit -- that of the sequential method. In this work, the basic idea is to improve the SNR of the 3pt correlator of SSM using the low lying eigenvectors of 
$D_c$ to construct the long distance part of the all-to-all  $S(\vec{y},t_2,\vec{z},t_1)$ {($S_L$ in Fig.~\ref{fig:quark_diagram}, the single line from the current to the sink)}, and using the noise many-to-all propagator to estimate the remaining high frequency part of $S(\vec{y},t_2,\vec{z},t_1)$ {($S_H$ in Fig.~\ref{fig:quark_diagram}, the double line from the current to the sink)}. Thus, the propagator
with LMS is written as
\bea    \label{eq:LMS_S}
S^{LMS_S}(\vec{y}_1,t_2,\vec{z},t_1)&=&\sum_i\theta^{(i)}_{\vec{y}_1} \gamma_5(S^{(i),H}_{noi}(\vec{z},t_1,t_2))^\dagger\gamma_5\nonumber\\
&&\qqe+\sum_i\frac{1}{\lambda_i + m} v_i(\vec{y},t_2)v_i^{\dagger}(\vec{z},t_1).
\eea 
where $\lambda_i$ and $v_i$ are the low-lying eigenvalues and the corresponding eigenvectors of $D_c$.
In other words, it is a technique to apply LMS to the sequential propagator $S_{seq}(X_{u,d};\vec{z},t_1,t_2,\vec{0},0)$ (LMS$_S$). It is expected to reduce the number of the noise propagators needed to reach the upper limit
{of SNR}.

When LMS$_S$  in Eq.~(\ref{eq:LMS_S}) is applied to the PropCI  in
Eq.~(\ref{eq:3pt}), $\hat{\cal S}$ comes from $t=0$ to $t=t_2$ through $t=t_1$
\bea  \label{eq:prop_ci_lms}
\hat{\cal S}^{LMS_S}&&\ \qqe\ ({\cal O},t_1;\vec{y},t_2,t_1,\vec{0},0)=\nonumber\\
&=&\sum_{\vec{z}} S^{LMS_S}(\vec{y}_1,t_2,\vec{z},t_1){\cal O}(\vec{z},t_1)S(\vec{z},t_1,\vec{0},0)\nonumber\\
&=&\sum_{\vec{z}} \big(\sum_i\frac{1}{\lambda_i + m} v_i(\vec{y},t_2)v_i^{\dagger}(\vec{z},t_1)\nonumber\\
&&\ \ \ \ \ \  + \theta^{(i)}(\vec{y},t_2)\sum_{\vec{z},i} \gamma_5(S^{(i),H}_{noi}(\vec{z},t_1,t_2))^\dagger\gamma_5\big)\nonumber\\
&&{\cal O}(\vec{z},t_1)S(\vec{z},t_1,\vec{0},0),
\eea
{as shown in the shadowed area in Fig.~\ref{fig:quark_diagram}.}

Then one can construct 3pt with LMS by constructing the standard 2pt repeatedly (the projection matrix $\Gamma$ is suppressed for clarity),

\begin{widetext}
\bea
C_3^{LMS,u}(\hat{\cal S}, S) &=&C^3_{ker}(\hat{\cal S}^{H},\hat{\cal S}^L,S^H_{NG},S^L_{NG},S^H_{NG},S^L_{NG})+\sum_{x\in G} C^3_{ker}(\hat{\cal S}^{L(x)},\hat{\cal S}^H,\theta(x)S^L,S^H_{NG},\theta(x)S^L,S^H_{NG})+\nonumber\\
&&C^3_{ker}(S^H_{NG},S^L_{NG},S^H_{NG},S^L_{NG},\hat{\cal S}^{H},\hat{\cal S}^L)+\sum_{x\in G} C^3_{ker}(\theta(x)S^L,S^H_{NG},\theta(x)S^L,S^H_{NG},\hat{\cal S}^{L(x)},\hat{\cal S}^H)\nonumber\\
C_3^{LMS,d}(\hat{\cal S}, S) &=&C^3_{ker}(S^H_{NG},S^L_{NG},\hat{\cal S}^{H},\hat{\cal S}^L,S^H_{NG},S^L_{NG})+\sum_{x\in G} C^3_{ker}(\theta(x)S^L,S^H_{NG},\hat{\cal S}^{L(x)},\hat{\cal S}^H,\theta(x)S^L,S^H_{NG})
\eea
\end{widetext}
where 
\bea   \label{eq:lms_3pt_ker}
S^L_{NG}&=&\sum_{x\in G} \theta(x)S^L(x),\textrm{ and}\nonumber\\
C^3_{ker}&&\qe\qe\qe(X_1,X_2,Y_1,Y_2,Z_1,Z_2)=  C(X_1,Y_1,Z_1)  \nonumber\\
&& + C(X_2,Y_1,Z_1)+ C(X_1,Y_2,Z_1) + C(X_1,Y_1,Z_2) \nonumber\\
\eea
and {$\hat{\cal S}^{H}$ and $\hat{\cal S}^{L(x)}$ are the high- and low-mode parts of  $\hat{\cal S}^{LMS_S}$
 in Eq.~(\ref{eq:prop_ci_lms}). }

This is the stochastic sandwich method with LMS which uses the low eigenmodes for
the propagator from the current to the sink in PropCI, $\hat{\cal S}^{LMS_S}$ with current insertion and the high modes for the same which originates from the sink time slice.
The construction of the PropCI with low modes needs to be done for each current and momentum transfer and $t_2$ (if desired). 
 In contrast, the current-sequential method will need to do an inversion for
each current, momentum transfer, and $t_1$ separately.

To account for the amount of numerical work for different approaches to the 3pt CI correlators, we 
note the the traditional sink-sequential method entails $2 \times 4 \times N_p$ inversions at a fixed sink time
slice $t_2$, where the 2 and 4 refer to the separate sources $X$ in Eq.~(\ref{eq:X_etc}) labeled with $u$ and $d$ flavors and polarization directions (unpolarized and polarization in 3 spatial directions). $N_p$ is the number
of sink momenta for the nucleon.  For SSM without LMS, there are $N_{noi}$ inversions of the $N_{noi}$ noise
vectors at the sink time $t_2$. How many $N_{noi}$ is needed for acceptable SNR depends on the observable. 
For the SSM with LMS, besides the noise propagator $S^{H}_{noi}$ with $N_{noi}^H$ inversion, there is 
an overhead for the low-mode portion of PropCI ($\hat{\cal S}^{LMS_S}$ in Eq.~(\ref{eq:prop_ci_lms})). It includes
$N$ times the low-mode contributions from $N$ smeared grid source plus one high-mode contribution for the
propagator from the source to the current ($S_{NG}^H$). Each needs to be folded with the current for different momentum transfer $\vec{q}$. Therefore the overhead is $\epsilon \times (N+1) \times N_{cu} \times N_q$ where $N_{cu}/N_q$ is the number of currents/momentum transfer, and $\epsilon$ is the fraction of inversion time for constructing the low-mode portion of $\hat{\cal S}^{LMS_S}$ for each current and momentum transfer. We list the cost  for the sink and current parts of the 3pt function in units of quark inversion in Table~\ref{table:cost} for future reference. To evaluate the efficacy among the three methods, one needs to compare costs in the table to reach
the same precision for a given observable. For the case of SSM with LMS, there is an additional gain from
the noise grid source with LMS as discussed in Sec.~\ref{sec:lms} which needs to be taken into account.

\begin{table}[htbp]    
\begin{center}
\caption{\label{table:cost}The cost for the sink and current parts of the 3pt function in units of quark inversion is
listed for the sink-sequential method (Sequential), stochastic sandwich method (SSM), and SSM with LMS. $N_p$ is the number of sink
nucleon momenta, $N_{noi}$ is the number of noise in SSM. $N_{noi}^H$ is the number of noise in 
SSM with LMS, and $N_{cu}/N_q$ is the number of currents/momentum transfer in the construction of of
the low-mode part of PropCI. $\epsilon$ is the fraction of inversion time for constructing the low-mode portion of PropCI for each current and momentum transfer and $N_p$ momenta ($\sim 0.02$ on the ensemble used in this work).}
\begin{ruledtabular}
\begin{tabular}{cccc}
Sequential & SSM & SSM+LMS$_S$\\
\hline
8$N_p$ & $N_{noi}$ &  $N_{noi}^H$+$\epsilon (N+1) N_{cu}N_q$
\end{tabular}
\end{ruledtabular}
\end{center}
\end{table}

\section{Numerical details}   \label{sec:numerical}

In this work, we use the valence overlap fermion on $2 +1$ flavor domain-wall fermion (DWF) configurations \cite{Aoki:2010dy} to carry out the calculation~\cite{Li:2010pw}. 

The lattice we use has a size $24^3\times 64$ with lattice spacing $a^{-1} = 1.75(4)$ GeV set by 
$r_0$ at the chiral and continuum limits~\cite{Yang:2014sea}. The light
sea $u/d$ quark mass $m_{l}a = 0.005$ corresponds to $m_{\pi} \sim 330$ MeV. We have calculated
the isovector matrix elements of the nucleon for the axial-vector and scalar couplings  and the quark momentum fraction at 6 valence quark mass parameters which correspond
to the renormalized masses $m_q^R\equiv m_q^{\overline{\rm MS}}(2 \rm GeV)$ ranging from 13 to 32 MeV after
the non-perturbative renormalization procedure in Ref.~\cite{Liu:2013yxz}. They correspond to the pion
mass in the range of 250-400 MeV.
In order to enhance the signal-to-noise ratio in the calculation of three-point functions, we use two
smeared noise 12-12-12 grid sources at $t_i=0$ and $32$ (one is PropNG and the one is PropNGM)~ \cite{Gong:2013vja} and two noise 2-2-2 grid point sources at positions $t_f$ which are 8, 10, and 12 
time-slices away from the sources on 203 configurations. 

The effective overlap operator $D_c$ is chiral, i.e. $\{D_c, \gamma_5\} = 0$ \cite{Chiu:1998gp}, and is expressed in terms of the overlap operator $D_{ov}$ as
\bea
D_c=\frac{\rho D_{ov}}{1-D_{ov}/2} \textrm{ with }D_{ov}=1+\gamma_5\epsilon(\gamma_5D_{\rm w}(\rho)),
\eea
where $\epsilon$ is the matrix sign function and $D_{\rm w}$ is the Wilson Dirac operator with a negative mass
characterized by the parameter $\rho=4-1/2\kappa$ for $\kappa_c < \kappa < 0.25$. We set $\kappa$=0.2  which corresponds to $\rho=1.5$.  

{
Compared to the earlier implementation of the overlap operator~\cite{Li:2010pw}, the current implementation further improves the performance of data exchange {on different nodes of the cluster} and uses the polynomial approximation for the overlap operator instead of the rational approximation, and has achieved better scaling and further speed up of the calculation by a factor of two on average~\cite{Alexandru:2011sc}.

The number of $D_c$'s low mode eigenvectors used for the deflation of the overlap operator inversion and LMS, on this $24^3\times 64$ lattice, is 200 pairs plus the zero modes, and the upper bound of the absolute value of the eigenvalues is 0.154 which is over two times larger than the dimensionless strange quark mass.
}

We check the efficacy of  the sequential low-mode substitution (LMS$_S$) in the PropCI by examining the 3pt functions for the isovector axial and scalar currents. We plot
the ratio of 3pt-to-2pt correlators as a function of the current insertion time $t_1$ in Fig.~\ref{fig:sink_seq} where  the sink time $t_2$ is 10. The blue dots and black triangles show the contributions  where the current-to-sink part of PropCI is from the low modes and the noise-estimated high modes respectively}. Notice that the contribution from the low modes is much larger than that of the high modes when the current time slice is farther away from the sink (i.e. closer to the source with small $t_1$) for both the axial and scalar cases, which reflects the fact that the low modes dominate the long-distance behavior of the {PropCI}  between $t_1$ to $t_2$. When the current is
closer to the sink with larger $t_1$, we see that the high modes dominate for the axial case which shows that
the high modes are important and dominate the short distance behavior of the propagator. However, the 
high-mode contribution is still small for the scalar current case when $t_1$ is close to the sink which shows that
the high-mode contribution is small for the 3pt function for the scalar current. 

The red squares are the sum of the low- and high-mode contributions from the present hybrid scheme. We have
also calculated the 3pt function without LMS$_S$ for the PropCI, but instead use only the noise propagator as
the full propagator from $t_1$ to $t_2$. These
are shown as the green triangles in Fig.~\ref{fig:sink_seq}. These correspond to the stochastic method
introduced by the QCDSF Collaboration \cite{Evans:2010tg,Bali:2013gxx} and the Cyprus group~\cite{Alexandrou:2013xon}. Since our LMS$_S$ replaces the long distance part of the current-to-sink part of {PropCI} with an {exact} all-to-all one, the larger its contribution the larger the improvement. As in Fig.~\ref{fig:sink_seq}, the blue dotd contribute
over 80\% in the $g_S^u$ case and so the improvement of LMS$_S$ is larger than in the $g_A^u$ case.
The error bars of SSM at the time slices $t_1 = 2 - 6$ turn out to be a factor $\sim 2$ for $g_A^u$ ($\sim 4$ for $g_S^u$) larger than that of using LMS$_S$ in the present approach.
\begin{figure}[tbh]
\begin{center}
\includegraphics[scale=0.7]{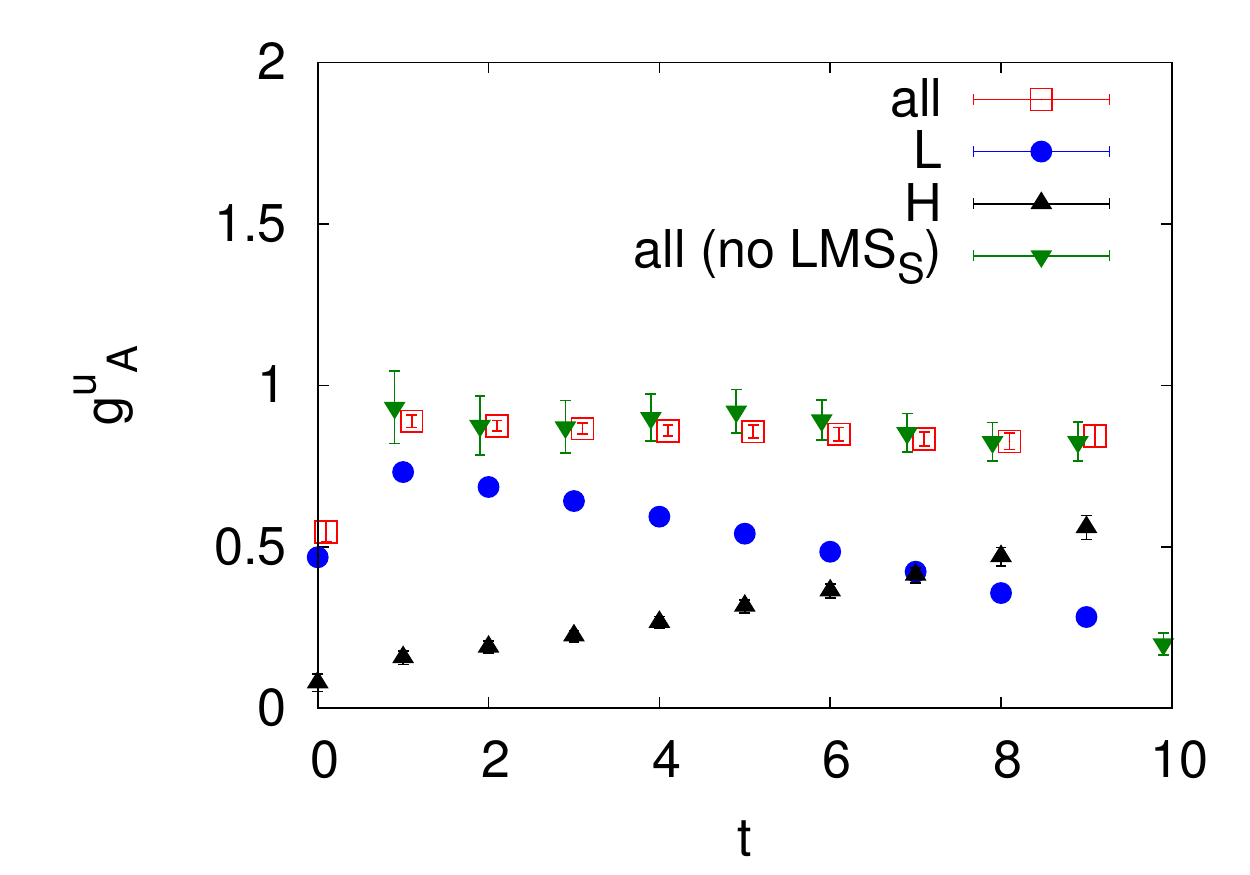}
  \includegraphics[scale=0.7]{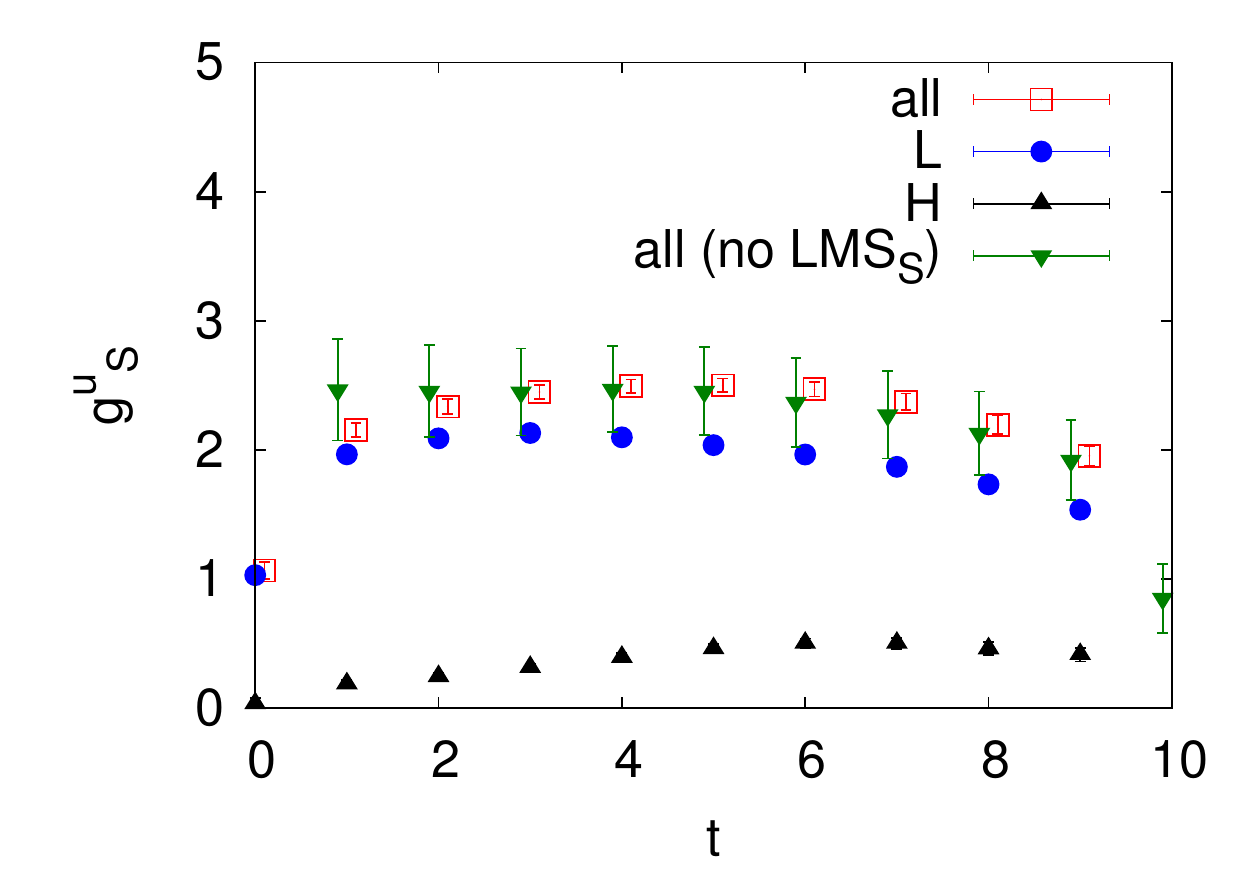}
\caption{\small The 3pt-to-2pt ratio with LMS$_S$ (red squares) vs. the one without it (green inverted triangles). The source/sink is  located at 0/10, and the current dependence for the matrix element with the current-to-sink part of PropCV including just the low- or high-mode parts are plotted as the blue dots or black triangles respectively. The upper panel is for the axial-vector current case and the lower panel is for the scalar case. Notice that the contribution of the low-mode part is larger when the current time slice is farther away from the sink. \label{fig:sink_seq}}
\end{center}
\end{figure}

\begin{figure}[tbh]
\begin{center}
  \includegraphics[scale=0.7]{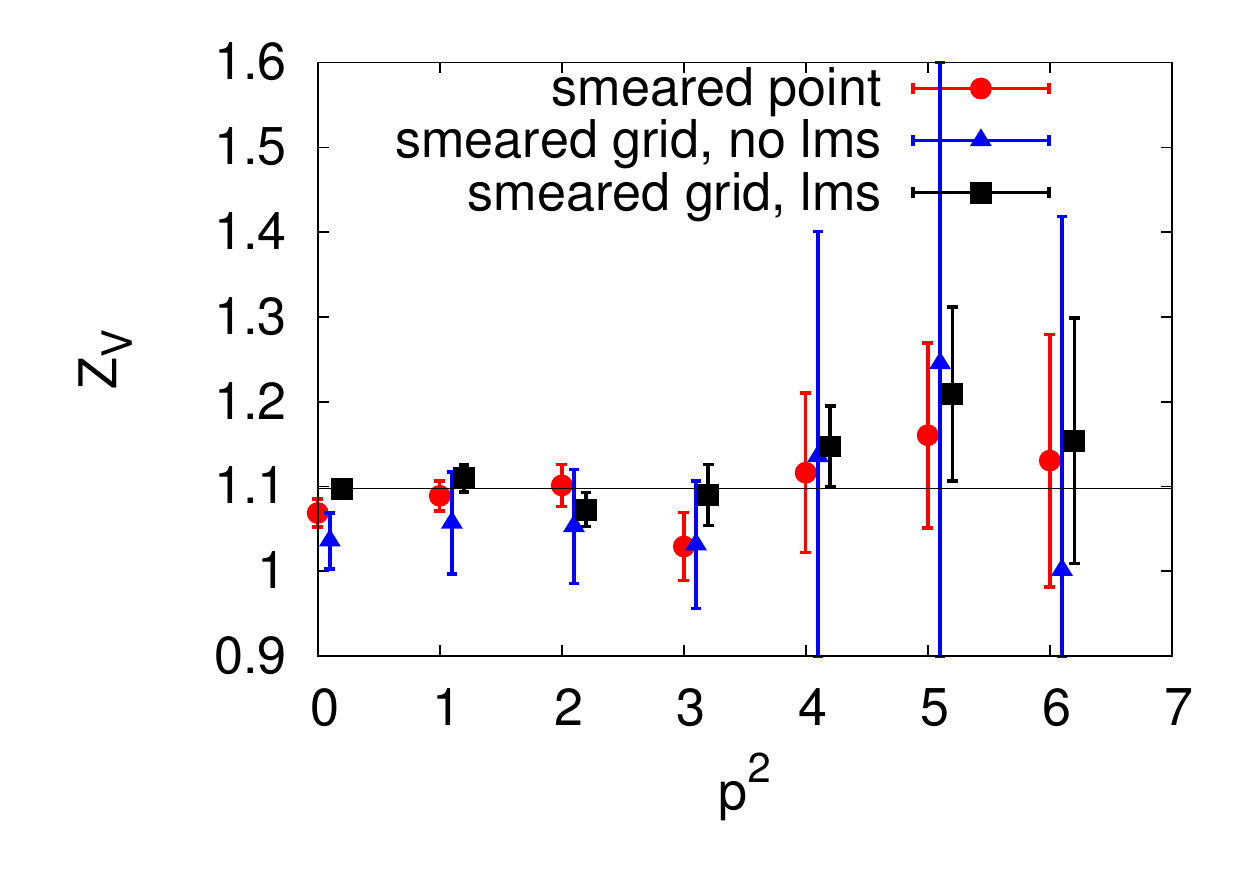}
\caption{\small The vector renormalization constant in the rest/moving frame at the unitary point, as a function of the momentum squared $p^2$ in lattice unit. The $p^2=0,4$ involve PropNG only and the other cases involve PropNGM also. The former case gains more from LMS (black squares vs. red dots). The results obtained using the 8-point grid source without LMS applied are very noisy (blue triangles).\label{fig:all_lms}}
\end{center}
\end{figure}

 The fact that the error of $g_A^3/g_S^3$ in our approach is smaller than that of SSM with 2 noises by
a factor of $\sim 2/4$ shows that it would take 8/32 noise inversions for SSM to have the same error as the present
method with LMS. To compare the cost of SSM + LMS, we should take its overhead into account. On the present lattice, the percentage of inversion time 
for low-mode construction is $\epsilon = 0.02$. Therefore, the overhead \mbox{$\epsilon (N+1) N_{cu} N_q = 0.72$} for $N= 8$ (smeared grid source), $N_{cu} = 4$ to account for the scalar current and $A_i$ for 3 spatial directions and $N_q = 1$. Together with $N_{noi}^H =2$, the cost is 2.72 inversions. This means that,
to reach the same error, it would take SSM  2.9 and 11.8 times more inversions than SSM with LMS
for $g_A^3$ and $g_S^3$ respectively. Furthermore, the smeared grid source with
LMS has improved the statistics by a factor of 5.3 for $N =8$ for the 2pt function. This additional factor
of improvement is also expected for the 3pt function. 

To compare with the sink-sequential method, we assume
that our results have reached the SNR of that of the sink-sequential method. This is consistent with the fact that in the range $t_1 = 2 - 6$ where the observables are fitted, the PropCI are dominated by the low-mode contributions, particularly for  $g_S^3$. In this case, the cost of sink-sequential takes 16 inversions. Here, we have taken $N_p =2$ to include the $\langle x\rangle_{u-d}$ calculation in addition to $g_A^3$ and $g_S^3$. For the overhead in
SSM + LMS, the number of currents needed is $N_{cu} = 6$ for these three quantities and the overhead is
$\epsilon (N+1) N_{cu} N_q = 1.08$. Therefore, besides the improvement from use of the grid source, the
present method would be $16/3.08 = 5.2$ times more efficient than the sink-sequential method for the
calculation of the three quantities. {Note that the cost of the sink-sequential method has additional factors {that need to be taken into account, such as} $N_{mass}$ for different masses, and also $N$ when the necessary LMS is applied on the source of the sink-sequential propagator (as in Eq.~\ref{eq:X_etc}), so SSM is much cheaper than the sink-sequential method.}

{
When the physical volume is increased, while keeping the lattice spacing unchanged, and with a noise vector covering the entire spatial volume of the sink time slice, we expect that the region essentially contributing to 3pt will not change, while the remaining region contributes only to the noise. Such a simple argument hints that the noise required to reach the same SNR is proportional to volume and we have confirmed it explicitly on the $48^3\times 96$ lattice with similar lattice spacing \cite{Blum:2014tka}. At the same time, the number of low modes will be proportional to volume if we want to reach the same upper bound of the eigenvalues, so the SSM with LMS will not lose its efficiency as 
compared to SSM without LMS, when the volume is larger. But, since the number of inversions is fixed in the standard sequential method, the SSM with and without LMS will lose their comparative efficiencies when the volume is very large.
}

Another issue we need to check is the effect of LMS in the 3pt case. For the 3pt function, we check, for example, the vector charge renormalization constant from the forward matrix element at the unitary point for several nucleon momenta. For  $p^2 = 0$ and 4, only the propagator PropNG is involved, while the other cases involve PropNGM also. In the former cases, we find that the smeared grid source with LMS improves the SNR by a factor of 2.0 compared to that with a smeared point source without LMS, slightly smaller than what we found with the 2pt function as discussed in Sec.~\ref{sec:lms}; whereas, the gain is only 1.4 for the other $p^2$ where the PropNGM is involved. 
We shall look into the possibility of improving the SNR further when PropNGM is  involved. 

\section{Results}\label{sec:res}

A standard 3pt/2pt ratio in the forward matrix element case is
\bea
&&\qqe R(t_{2},t_1, 0)=C_3(t_2,t_1,0)/C(t_2,0)\nonumber\\
&=&\frac{\sum_{i,j} Z^{(i)}_{\textbf{f}}Z^{(j)}_{\textbf{i}}e^{-E^{(i)}(t_{2}-t_1)-E^{(j)}t_1}\langle \chi^{(i)}_{\textbf{f}}|J|\chi^{(j)}_{\textbf{i}}\rangle}{\sum_{k} Z^{(k)}_{\textbf{f}}Z^{(k)}_{\textbf{i}}e^{-E^{(k)}t_{2}}}\nonumber\\
& _{\overrightarrow{t_2\gg0}}&\langle \chi^{(0)}_{\textbf{f}}|J|\chi^{(0)}_{\textbf{i}}\rangle\nonumber\\
&&
+\frac{Z_f^{(1)}}{Z_f^{(0)}}\langle \chi^{(1)}_{\textbf{f}}|J|\chi^{(0)}_{\textbf{i}}\rangle e^{-\Delta E(t_{2}-t_1)} \nonumber\\
&&
+\frac{Z_i^{(1)}}{Z_i^{(0)}}\langle \chi^{(0)}_{\textbf{f}}|J|\chi^{(1)}_{\textbf{i}}\rangle e^{-\Delta E t_1}\nonumber\\
&&
+\frac{Z_f^{(1)}Z_i^{(1)}}{Z_f^{(0)}Z_i^{(0)}}(\langle \chi^{(1)}_{\textbf{f}}|J|\chi^{(1)}_{\textbf{i}}\rangle-\langle \chi^{(0)}_{\textbf{f}}|J|\chi^{(0)}_{\textbf{i}}\rangle
) e^{-\Delta E t_{2}}\nonumber\\
&&
+...,\label{eq:excited_state}
\eea
where $E^{(i)}$ and $Z^{(i)}$ are the energy and the overlap of the interpolation field of the $i$th state and 
$\Delta E=E^{(1)}-E^{(0)}$. For $t_{2}\gg t_1\gg0$, the contributions from all the terms in the right hand of Eq.~(\ref{eq:excited_state}) except the first term vanish, and then one can use Eq.~(\ref{eq:excited_state}) to obtain the matrix element. 

When $t_{2}$ is fixed, one may fit the first term and the combined second and third terms around 
$t_1 = t_{2}/2$ to include the effect of the ground state to first excited state transition in the right hand side of Eq.~(\ref{eq:excited_state}) which is $t_1$ dependent. But since the fourth term in the right hand side of Eq.~(\ref{eq:excited_state}), which is the difference of the matrix element in the ground state and the first excited state, is independent of $t_1$ just like the first term, one will not be able disentangle them and, as a result, a systematic error may be induced by its contribution which is suppressed by $e^{-(E^{(1)}-E^{(0)})t_{2}}$. To get a feeling for the size of the correction, let us suppose that the first excited state matrix element 
$\langle \chi^{(1)}_{\textbf{f}}|J|\chi^{(1)}_{\textbf{i}}\rangle$ is 30\% different from the ground state 
matrix element $\langle \chi^{(0)}_{\textbf{f}}|J|\chi^{(0)}_{\textbf{i}}\rangle$, and the mass difference of the first excited state and the ground state is about 500 MeV. Then the correction from such a effect with $t_{2}$=8, 10 and 12 (with the nucleon source set at $t_0 =0$) is about 3\%, 2\% and 1\% respectively. To assess this error, we shall calculate
the 3pt function at three values of $t_2$ so that we can fit all four terms in Eq.~(\ref{eq:excited_state}).

\begin{figure}[tbh]
\begin{center}
  \includegraphics[scale=0.7]{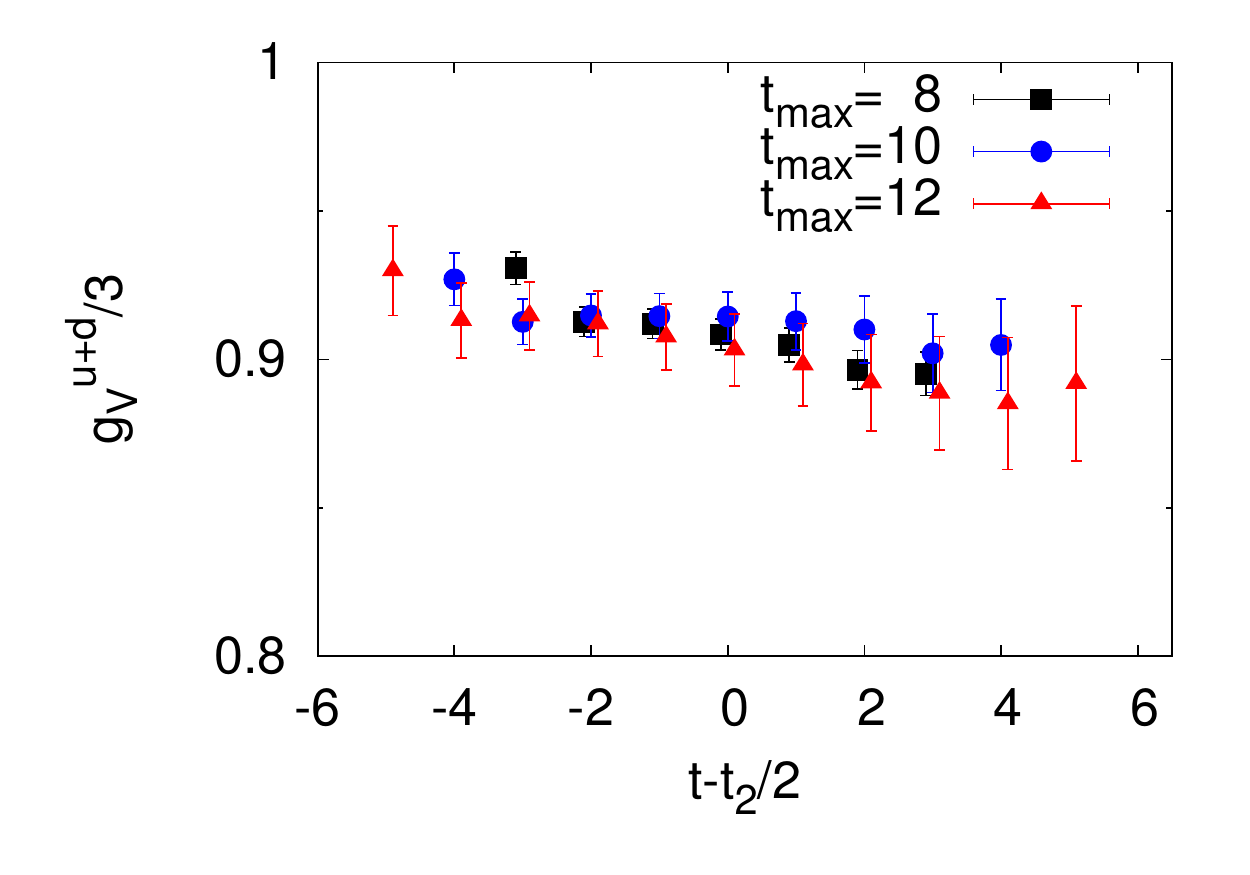}
\caption{\small The nucleon sink-source separation dependence of the matrix element with the vector charge for $u +d$ in the connected insertion. Obviously, the larger $t_{2}$, the worse the signal. The data points marked with the black squares ($t_{2}$=8), the blue dots ($t_{2}$=10) and the red triangles ($t_{2}$=12) are consistent. \label{fig:tdep_vec}}
\end{center}
\end{figure}

In order to check the 
$t_{2}$ dependence of the plateau, three sets of propagators with two noise-grid point sources 
each at positions $t_2 =8, 10$ and 12 time-slices away from the nucleon source are generated, and all the $t_1$ dependence of these three cases are plotted together for comparison in Fig.~\ref{fig:tdep_vec} for the vector current case. The sink-source separation dependence seems to be mild here, but in general the minimum separation required by other quantities can be different.

To check the separation effect quantitatively, we applied three kinds of fits to deduce the results:

{The first method is to fit the ratio as a function of $t_1$ and $t_{2}$,
\bea
&&R^{fit}(t_{2},t_1)=C_0+C_1e^{-\Delta m(t_{2}-t_1)}\nonumber\\
&&+C_2e^{-\Delta m t_1}+C_3e^{-\Delta m t_{2}}
\eea
with $C_{0,1,2,3}$ and $\Delta m$ as free parameters. $C_0$ is the ground state matrix element we want. Since the $t_1$ dependence of $R(t_{2},t_1)$ is mild in some of the quantities like $g_V$ and $g_A$, we take $\Delta m$ as a common parameter for all the quantities. This is what we mark as ``2-state" in the following discussion.

In this work, we use the smeared source and the point sink, so the excited-state contaminations are different in the smaller and larger $t$ ends. If the smeared source makes the contaminations in the smaller $t$ end small, or has a different sign compared to that in the larger $t$ end, the position of the plateau will be harder to determine, as in the case of $g_V^{u+d}$ (Fig. \ref{fig:tdep_vec}) and $g_A^3$ (Fig. \ref{fig:ga1}).
Applying the ``2-state" fit on such a quantity is not stable and provides large uncertainties (and/or large $\chi^2/d.o.f.$) on the results. In this work, we constrain the mass difference $\Delta m$ to be the same for the different matrix elements with the same quark mass value, and apply a correlated joint 2-state fit. To suppress the contamination from the excited state, we excluded the data points with $t_1=0,t_2-1$ and $t_2$. One more data point at the larger $t$ end is excluded since the excited-state contamination is larger there. Despite this, the fit is still not very good. Taking the unitary point as an example, the $\chi^2/d.o.f.$ with $\sim$70 degrees of freedom is 1.45, the corresponding p-value is just 0.008. In addition, this method requires a joint fit with several quantities and is not suitable for the analysis of a single quantity without the information of the other quantities.

The second method is the sum method  \cite{Maiani:1987by,Deka:2008xr} which is used in the disconnected insertion case, wherein a sum is taken over all the 3pt/2pt ratios in Eq.~(\ref{eq:excited_state}) with different $t_1$, 
\bea
&&\qqe SR(t_{2},t_1,0)=\sum_{0 < t_1 < t_{2}} R(t_{2},t_1,0)\nonumber\\
&&=(t_{2}-1)\langle \chi^{(0)}_{\textbf{f}}|J|\chi^{(0)}_{\textbf{i}}\rangle\nonumber\\
&&+\frac{e^{-\Delta m}}{1-e^{-\Delta m}}(\frac{Z_f^{(1)}}{Z_f^{(0)}}\langle \chi^{(1)}_{\textbf{f}}|J|\chi^{(0)}_{\textbf{i}}\rangle +\frac{Z_i^{(1)}}{Z_i^{(0)}}\langle \chi^{(0)}_{\textbf{f}}|J|\chi^{(1)}_{\textbf{i}}\rangle)\nonumber\\
&&+(t_{2}-1)\frac{Z_f^{(1)}Z_i^{(1)}}{Z_f^{(0)}Z_i^{(0)}}(\langle \chi^{(1)}_{\textbf{f}}|J|\chi^{(1)}_{\textbf{i}}\rangle-\langle \chi^{(0)}_{\textbf{f}}|J|\chi^{(0)}_{\textbf{i}}\rangle)\nonumber\\
&&\ \ \ \  e^{-\Delta m t_{2}}+...
\eea

When $t_{2}$ is large, we can use the linear function of $t_{2}$ (ignoring the $e^{-\Delta m t_{2}} $ correction)
\bea
 SR^{fit}(t_{2},t_1)&=&t_{2} C_0+C'_1
\eea
to fit our summed ratio with 3 different separations, and obtain the slope as the ground state matrix element. This method will be marked as ``sum" in the following discussion. 

We found that the ``sum" fit can obtain a $\chi^2/d.o.f.$ smaller than one, for all the quantities. But this fit just has one degree of freedom. Ignoring the $e^{-\Delta m t_{2}}$ correction can induce an uncontrolled systematic error.

The third method is to combine the first two methods, by fitting both the ratios and their sum together (denoted as ``mixed"),
\bea
R^{fit}(t_{2},t_1)&=&C_0+C_1e^{-\Delta m(t_{2}-t_1)}\nonumber\\
&&+C_2e^{-\Delta m t_1}+C_3e^{-\Delta m t_{2}},\\
 SR^{fit}(t_{2},t_1)&=&t_{2} C_0+\frac{e^{-\Delta m}}{1-e^{-\Delta m}}(C_1+C_2)\nonumber\\
&&+(t_{2}-1)C_3 e^{-\Delta m t_{2}} +C_4,
\eea
where $C_{0,1,2,3}$ and $\Delta m$ are the same as that in the ``2-state" fit, and $C_4$ is for the constant contribution from the transition between higher excited states and the ground state. 

The ``2-state" fit makes fully use of the ratios, while it is unstable when the position of the plateau is hard to determine (such as for $g^3_A$). The ``sum" fit provides a stable estimate of the ground state matrix element, but it suffers from the systematic error from ignoring the $e^{-\Delta m t_{2}}$ correction. By combining them together, we can obtain a stable fit of all the quantities discussed in this work independently, and don't have to use a joint fit with several quantities. The $\chi^2/d.o.f.$ of different quantities and quark masses vary between 1.0 and 1.5 with 18 degrees of freedom, corresponding to p-values in the range of [0.08-0.46]. The value of $\Delta m$ we obtained at the unitary point has a strong dependence on the quantity and varies from 400 MeV to 1GeV. 
}

The values for the renormalized isovector axial vector coupling $g^3_A$, scalar coupling $g^3_S$ and quark momentum fraction $\langle x\rangle_{u-d}$ from the three methods at the unitary point are listed in Table II.

\begin{table}[htbp]
\begin{center}
\caption{\label{table:axial}Isovector axial-vector coupling $g_A^3$, scalar coupling $g^3_S$ and quark momentum fraction $\langle x\rangle^E_{u-d}$ at the unitary point from three fitting methods. See the following three subsections for the details.}
\begin{ruledtabular}
\begin{tabular}{c|ccc}
& 2-state &sum & mixed \\
\hline
$g_A^3$ &  1.189(20) &1.157(18)& 1.166(19)\\
$g^3_S$&  0.61(6) &0.78(6) & 0.74(4)\\
$\langle x\rangle^E_{u-d}$ &  0.209(12) &0.190(13)& 0.193(19)
\end{tabular}
\end{ruledtabular}
\end{center}
\end{table}

\subsection{Vector and Axial vector case}

\begin{figure}[tbh]
\begin{center}
  \includegraphics[scale=0.7]{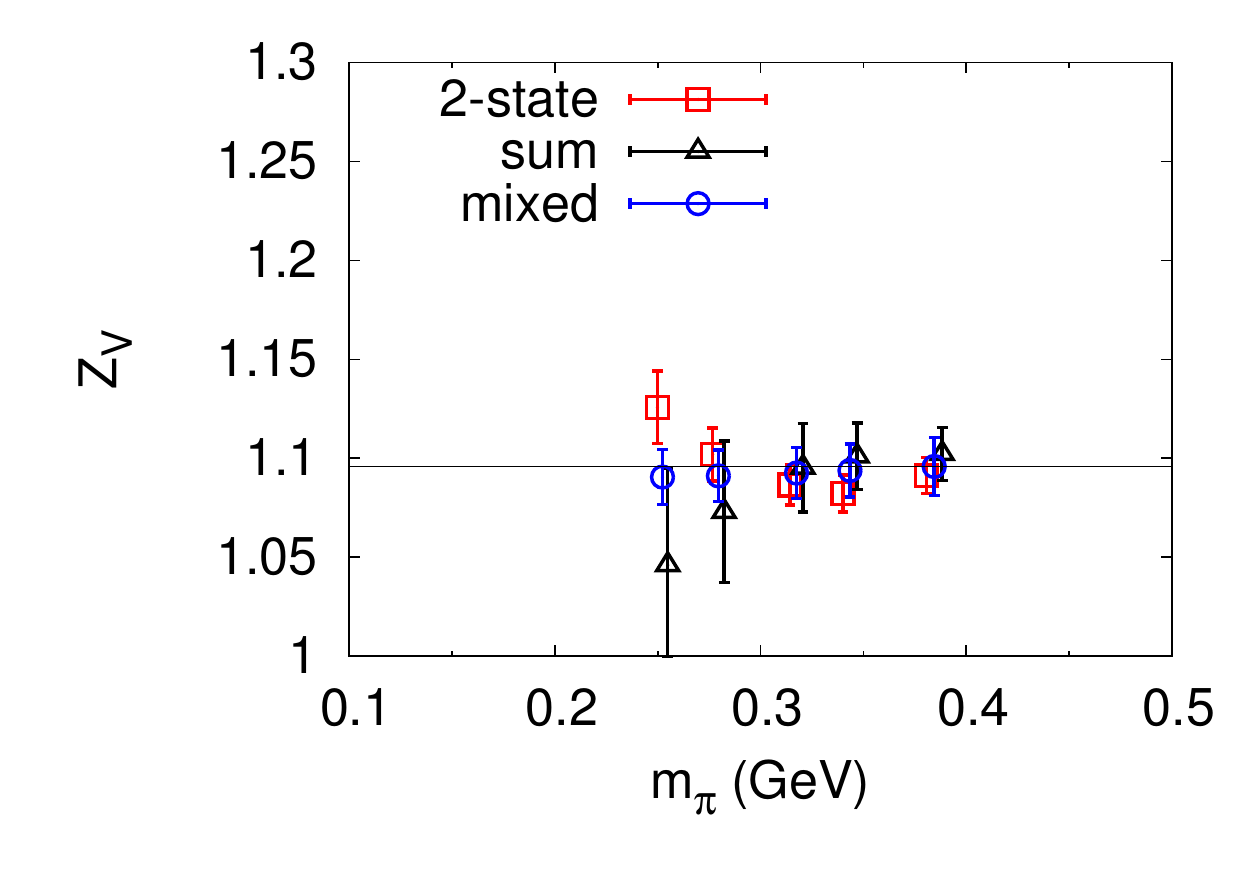}
\caption{\small The vector renormalization factor from the charge vs the pion mass, from three kinds of fitting methods: 2-state fit (red squares), summed slope (black triangles), and the mixed fit which combines those two methods (blue dots). The results from these different methods are consistent while that from the mixed method provides the best signal. \label{fig:gv}}
\end{center}
\end{figure}

The lattice renormalization of the vector current can be defined from normalizing the vector charge, 
\bea
g_{V_{4}}^{b}&\equiv&\frac{\textrm{Tr}[\Gamma^e\langle P |\int d^3x \overline{\psi}(x)\gamma_{4}\psi(x) |P \rangle]}{\textrm{Tr}[\Gamma^e\langle P|P \rangle]}=\frac{1}{Z_V}
\eea
where  superscript $b$ is for bare value, and $\Gamma^e=(1+\gamma_4)/2$ is the unpolarized projection operator. Fig.~\ref{fig:gv} shows that all the fitting methods mentioned in the last section provide consistent results, while the results from the ``mixed" method have the best signals among the three methods. A constant fit for the cases with $m_{\pi}\in(0.25,0.4)$ GeV gives the value of the vector renormalization factor as 1.096(6) which is just slightly smaller than the value 1.105(4) obtained from the axial Ward identity \cite{Liu:2013yxz}. 

\begin{figure}[tbh]
\begin{center}
   \includegraphics[scale=0.7]{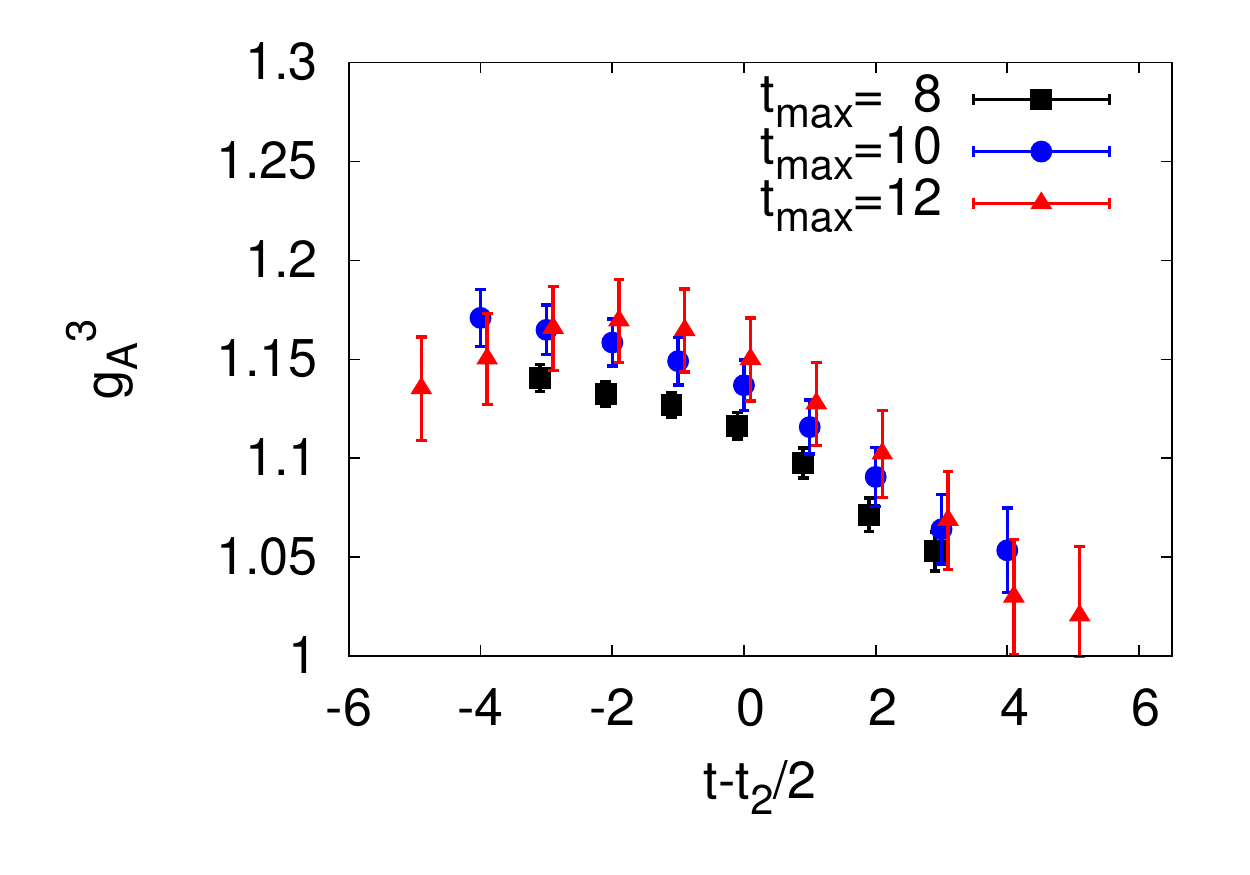}
\caption{\small The sink-source separation dependence of the matrix element of the isovector axial-vector current. The data of the isovector case with $t_{2}=8$ (black squares) are slightly smaller than that from the other two separations, while the result with $t_2$=10 (blue dots) is consistent with that with $t_2$ =12 (red triangles). \label{fig:ga1}}
\end{center}
\end{figure}

Then the renormalization of the vector current can be used to renormalize the axial-vector matrix element with polarized projection, 
\bea  
g_A^{b}&\equiv&\frac{\sum_{i=1,2,3}\textrm{Tr}[\Gamma^m_i\langle P |\int d^3x \overline{\psi}(x)\gamma_{5}\gamma_{i}\psi(x) |P \rangle]}{3\textrm{Tr}[\Gamma^e\langle P|P \rangle]}\nonumber\\
g_A^{{R}}&\equiv&g_A^{b}Z_V\nonumber\\
&=&\frac{\sum_{i=1,2,3}\textrm{Tr}[\Gamma^i\langle P |\int d^3x \overline{\psi}(x)\gamma_{5}
{\gamma_i}\psi(x) |P \rangle]}{3\textrm{Tr}[\Gamma^e\langle P |\int d^3x \overline{\psi}(x)\gamma_{4}\psi(x) |P \rangle]} \label{renorm_gA}
\eea
where the superscript $b/R$ stands for the bare/renormalized value respectively and
$\Gamma^m_i=(1+\gamma_4)\gamma_i\gamma_5/2$ is the polarized projection operator. 

\begin{figure}[tbh]
\begin{center}
  \includegraphics[scale=0.7]{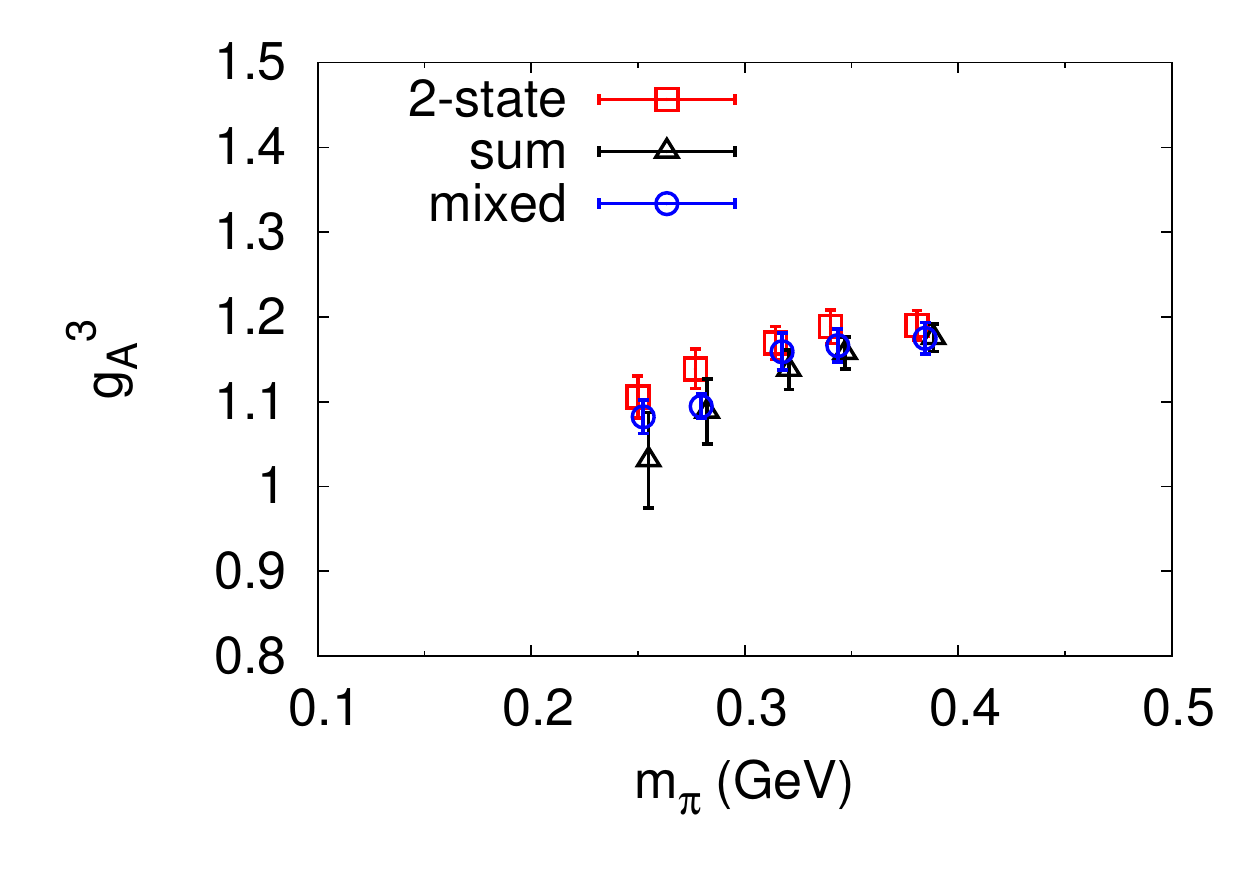}
\caption{\small The isovector axial-vector matrix element vs the pion mass, from three kinds of fitting method: 2-state fit (red squares), summed slope (black triangles), and the mixed fit which combines those two methods (the blue dots). The results from these different methods are consistent while that from the mixed method provides the best signal. \label{fig:ga2}}
\end{center}
\end{figure}

Using $g_{V_4}^{b}$ (instead of that from the axial Ward identity for pion) to renormalize 
$g_A$ as in Eq.~(\ref{renorm_gA}) could improve the signal of the renormalized $g_A$ by $\sim$20\% since these two matrix elements are correlated. As observed in Fig.~\ref{fig:ga1}, the sink-source separation dependence for the isovector case is mild, while a curve is observable at the right side of the plateau due to a larger excited state contribution from the point interpolation field at the sink. This is in contrast to the flatter behavior to the left of the plateau where the excited-state contribution is ameliorated by the smeared source. In Fig.~\ref{fig:ga2}, we plot the results of the isovector axial-vector coupling $g_A^3$ from the three fitting methods we mentioned. We note that those from the {``mixed" method are always between those from the other two methods}, for all the data points in the range of $m_{\pi}\in(0.25,0.4)$ GeV. The values from the three methods at the unitary point are listed in Table \ref{table:axial}. Similar to other lattice calculations at this pion mass (i.e. $\sim 300$ MeV), irrespective of which fit is used, the isovector axial-vector matrix element, $g_A^{u-d}$ is $\sim$10\% smaller than the experimental value 1.2723(23)\cite{Agashe:2014kda}.

\subsection{Scalar case}

\begin{figure}[tbh]
\begin{center}
  \includegraphics[scale=0.7]{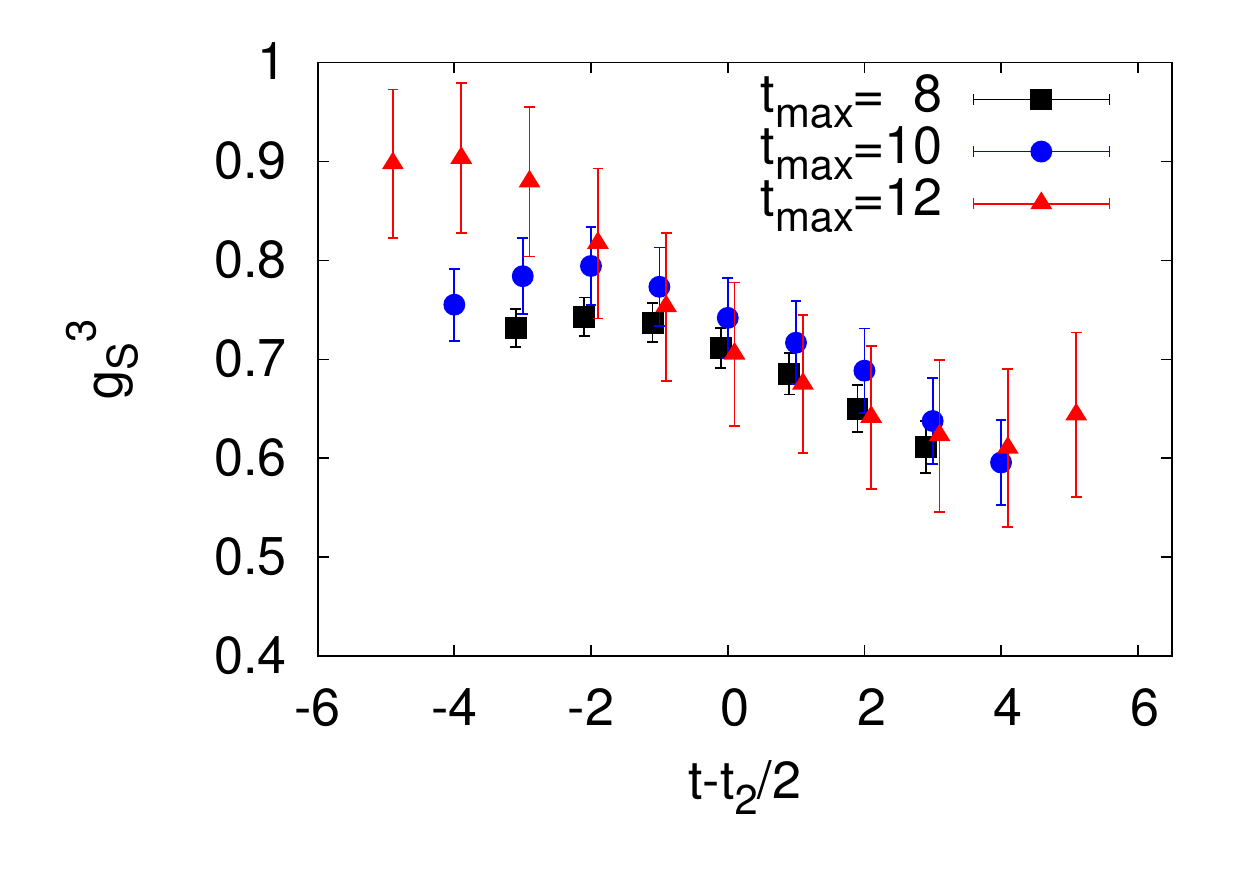}
  \includegraphics[scale=0.7]{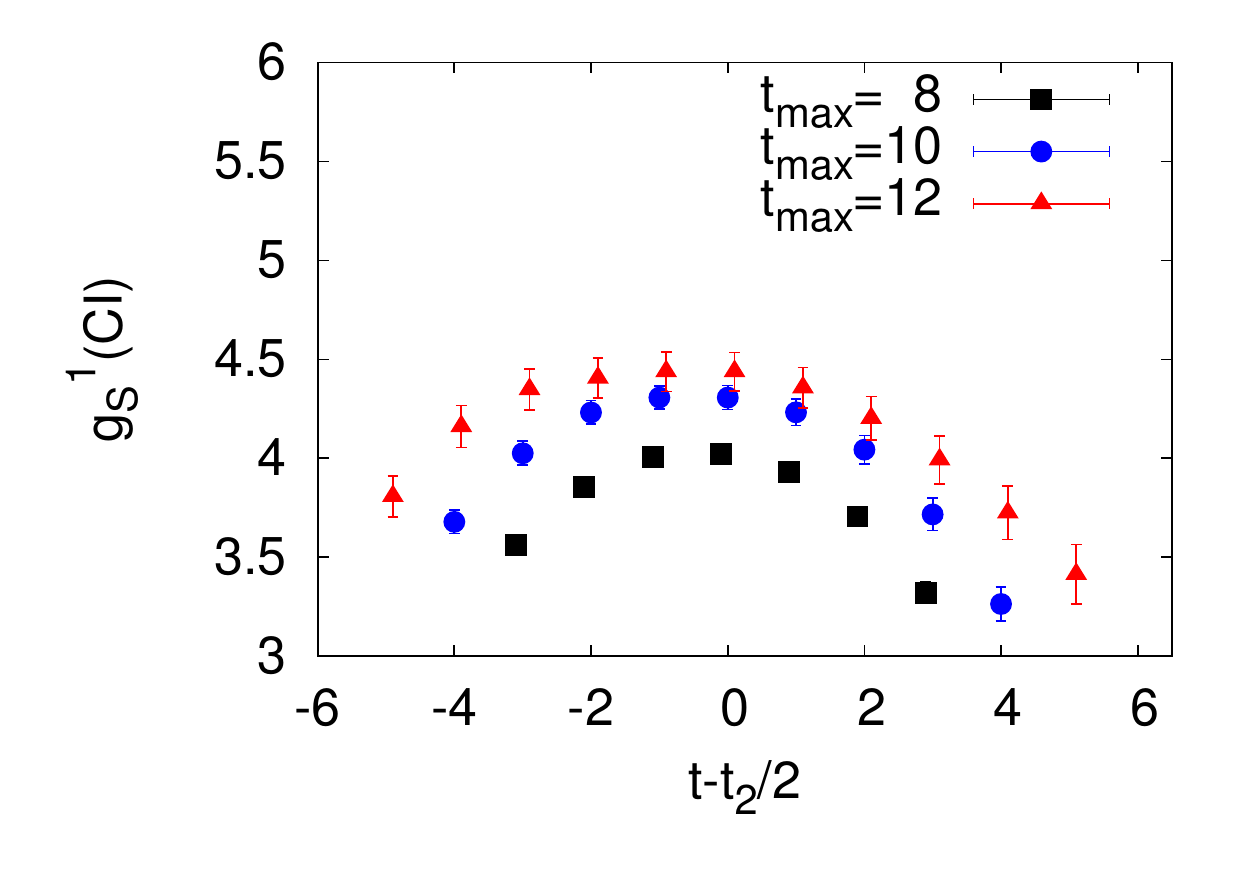}
\caption{\small The separation dependence of the matrix element of scalar current, for both the isovector and the CI part of the singlet case. The dependence is mild for the isovector case (the upper panel), while obvious for the CI part of the singlet case (the lower panel).\label{fig:gs1}}
\end{center}
\end{figure}

Similarly, the renormalized scalar matrix element with the unpolarized projection of the nucleon can be calculated by,
\bea
g_S&\equiv&\frac{Z_S\textrm{Tr}[\Gamma^e\langle P |\int d^3x \overline{\psi}(x)\psi(x) |P \rangle]}{\textrm{Tr}[\Gamma^e\langle P|P \rangle]},
\eea
where the renormalization constant $Z_S$ is obtained from the RI/MOM scheme and its value on the ensemble we use here is calculated to be 1.1397(54) \cite{Liu:2013yxz}. On the other hand, if one just focuses on the $\pi N\sigma$ term, $2Z_m m_q Z_S g_S^b$, the renormalizations of the quark mass $Z_m$ and that of the scalar matrix element $Z_S$ are canceled and so the $\pi N\sigma$ term is free of the renormalization. 

\begin{figure}[tbh]
\begin{center}
  \includegraphics[scale=0.7]{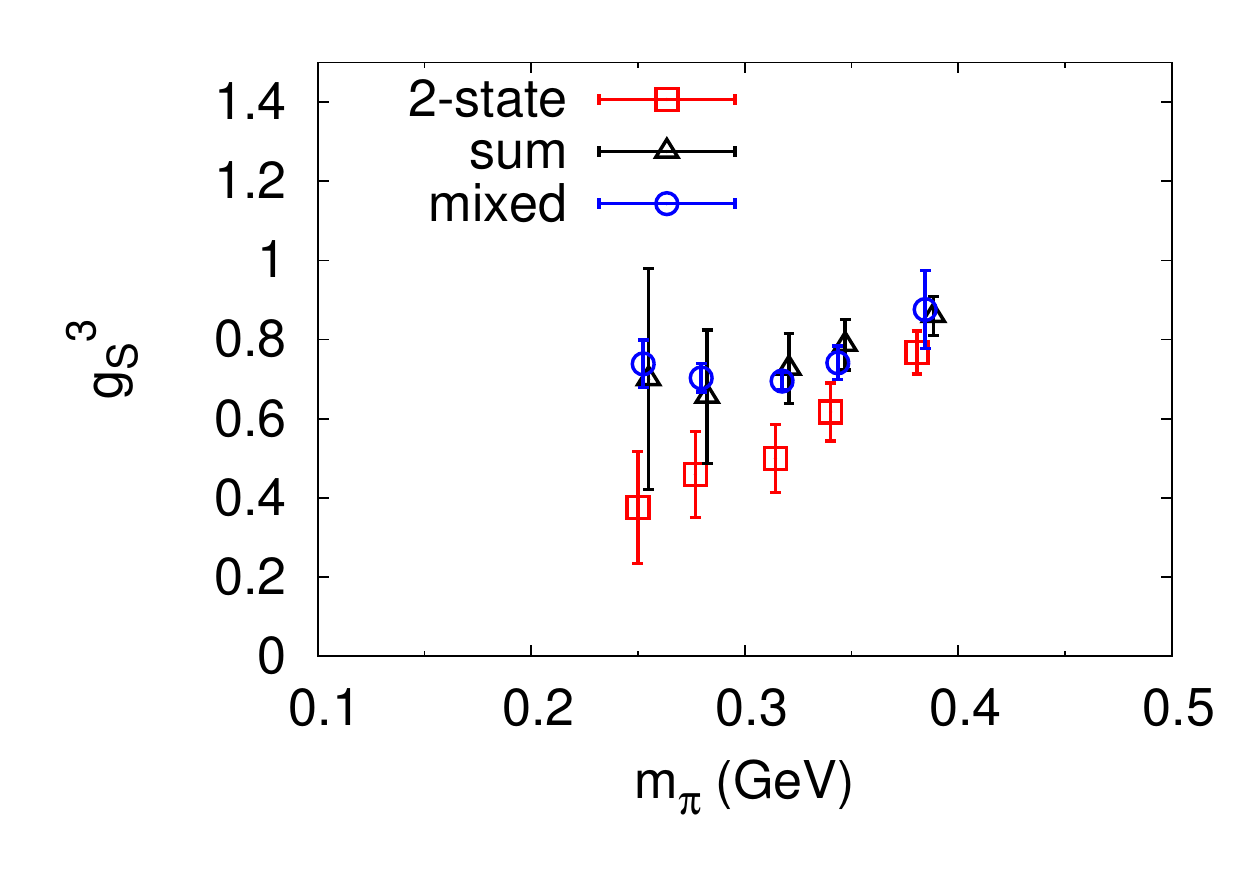}
\caption{\small Isovector scalar matrix element vs. pion mass from three kinds of fitting method: 2-state fit (red squares), summed slope (black triangles), and the mixed fit which combines those two methods (blue dots). The results from these different methods are slightly different.  \label{fig:gs2}}
\end{center}
\end{figure}

It is interesting to point out that the CI part of the scalar singlet matrix element has a strong 
sink-source separation dependence, as seen in the lower panel  of 
Fig.~\ref{fig:gs1}. At the same time, such a separation dependence seems to be canceled between the $u$ and $d$ quarks, so that the isovector case in the upper panel of Fig.~\ref{fig:gs1} has only a mild separation dependence. The results for the isovector scalar matrix element from the three fitting methods are plotted in Fig.~\ref{fig:gs2} and those at the unitary point are listed in 
Table \ref{table:axial}.
 This shows that, despite the fact that there are 2 $u$ valence quarks and only one $d$ quark in the proton, the $d$ contribution to the scalar matrix element per quark is more  than that of the $u$, as 
\bea
\frac{g_{S,CI}^u}{2g_{S,CI}^d}=0.67(2)
\eea
is much smaller than one. The scalar matrix elements of both the $u$ and $d$ quark increase as $m_q$ decreases, but the isovector scalar matrix element is not far from unity over the entire quark mass region from light to heavy. This has been interpreted to be
related to the Gottfried sum rule violation~\cite{Liu:1993cv} where it is found experimentally that there are more $d$ antipartons than $u$ antipartons..

\subsection{Quark momentum fraction}

\begin{figure}[!h]
\begin{center}
  \includegraphics[scale=0.7]{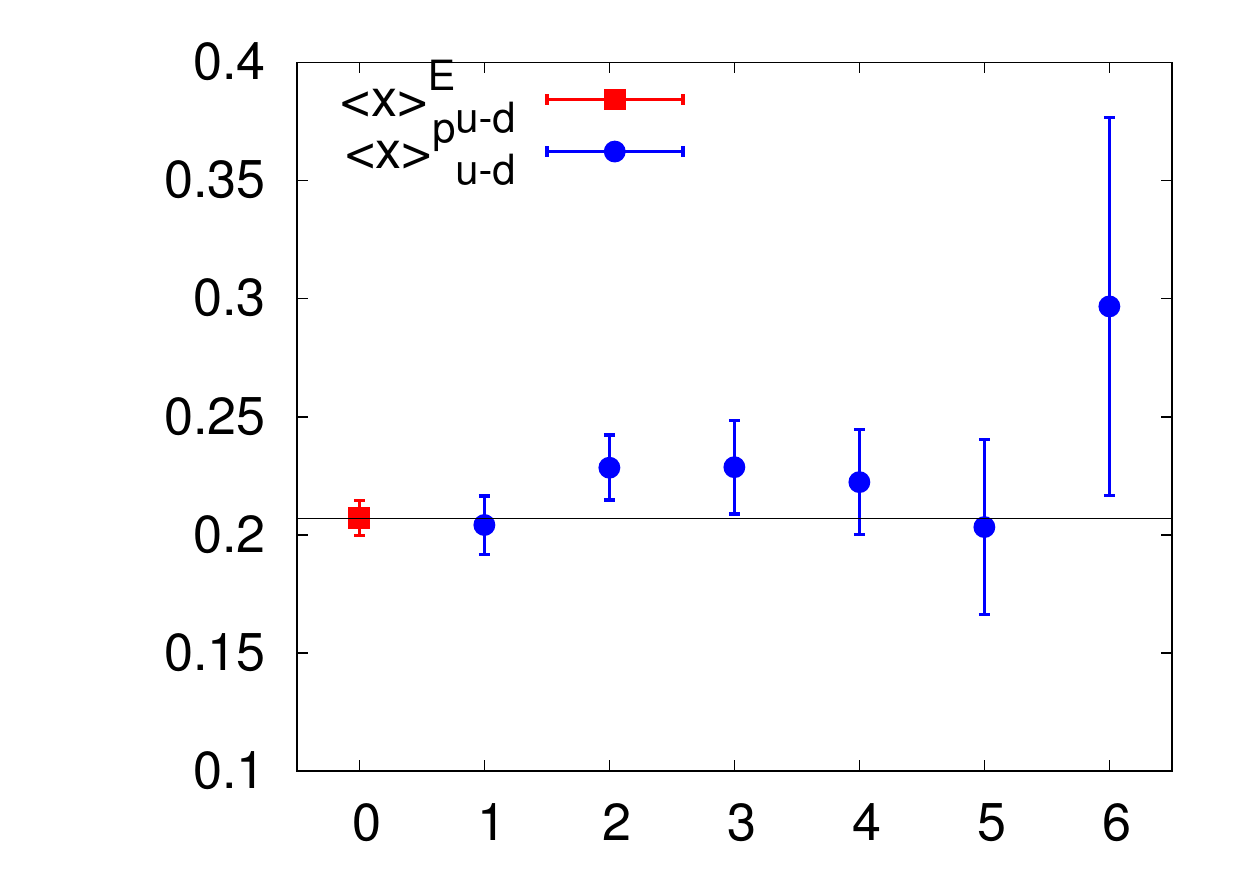}
\caption{\small The plateau fit values ($t_2 = 10$ case) of  the isovector momentum fraction for
$\langle x\rangle_{\textrm{E}}$ in the rest frame (red square) and also 
$\langle x\rangle_{\textrm{P}}$ in a moving frame with different momenta (blue dots). The results from both the diagonal and off-diagonal components (and also that from different momenta based on the off-diagonal matrix components) are consistent, but the first approach provides much better SNR. \label{fig:x0}}
\end{center}
\end{figure}

\begin{figure}[!h]
\begin{center}
  \includegraphics[scale=0.7]{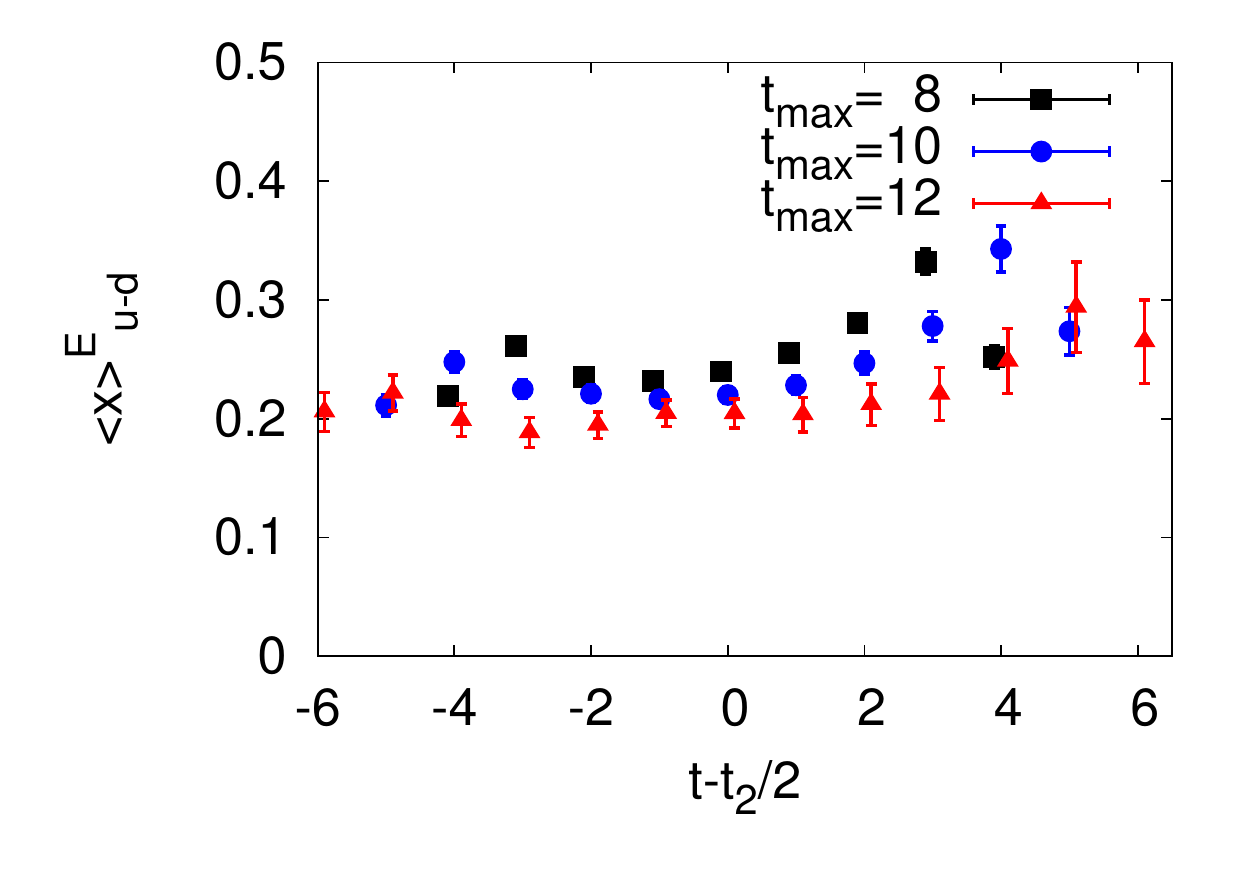}
  \includegraphics[scale=0.7]{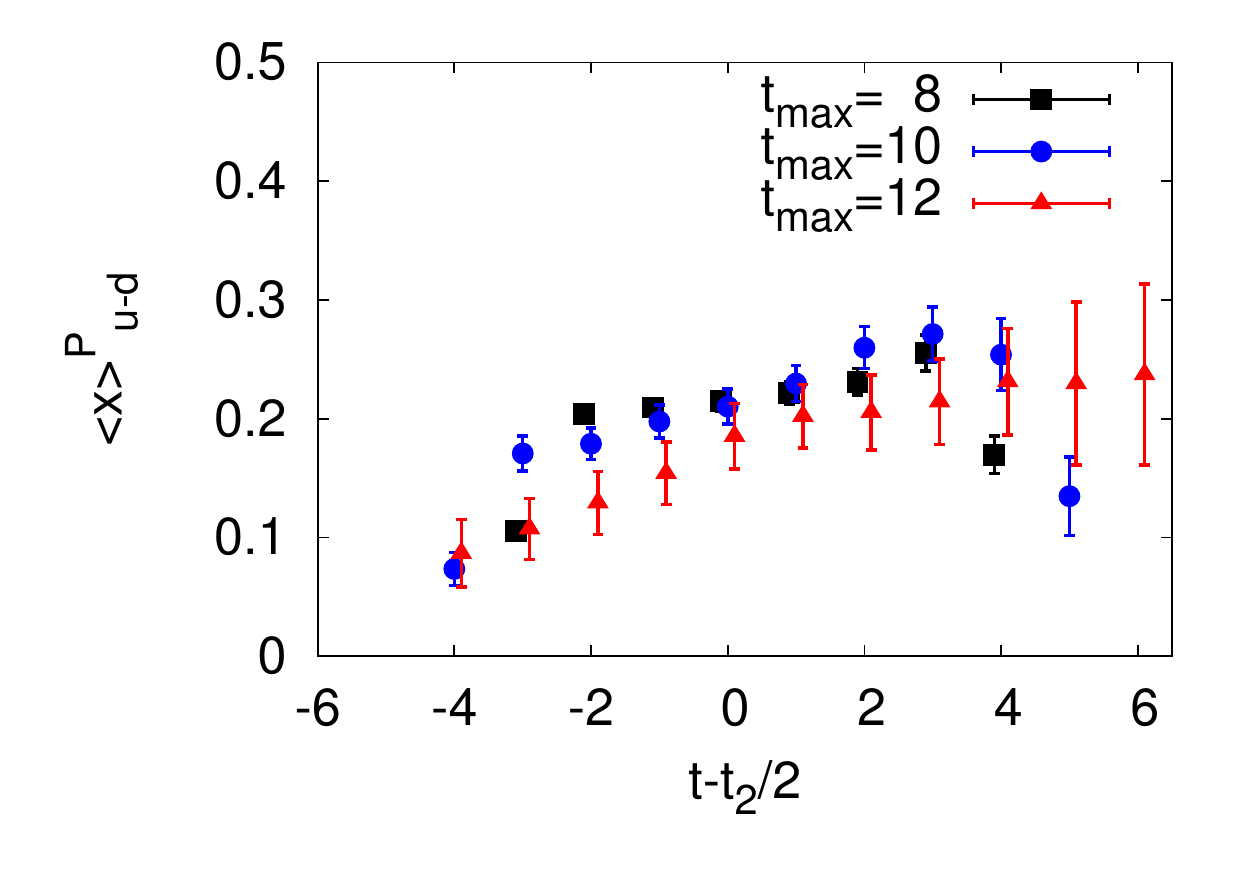}
\caption{\small The sink-source separation dependence of the isovector quark momentum fraction
for the case of the diagonal components of the energy momentum tensor (the upper panel) and that of the off-diagonal ones (the lower panel).\label{fig:x1}}
\end{center}
\end{figure}

\begin{figure}[!h]
\begin{center}
  \includegraphics[scale=0.7]{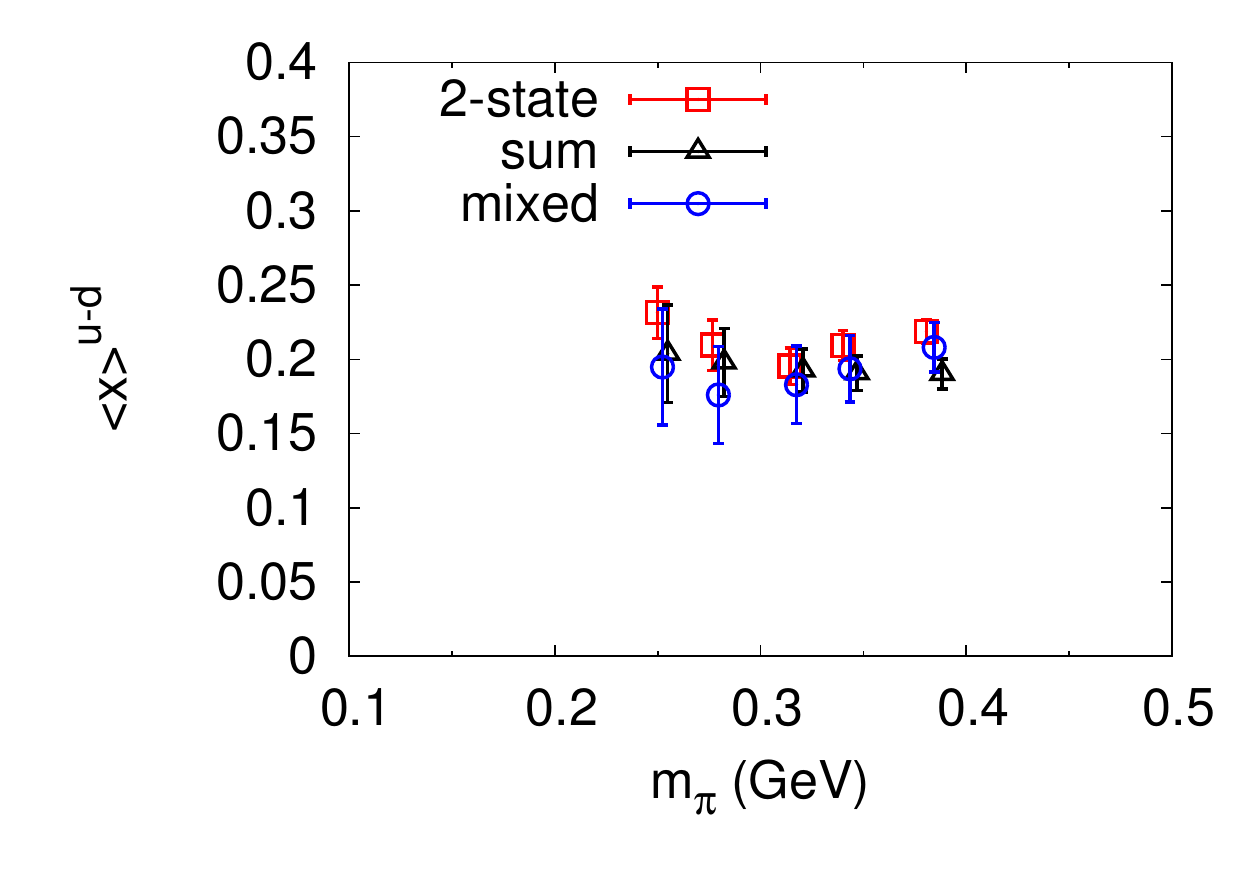}
\caption{\small The isovector quark momentum fraction ($\langle x\rangle^{\textrm{E}}$ in the rest frame) vs. the pion mass, from three fitting methods: 2-state fit (red squares), summed slope (black triangles), and the mixed fit which combines those two methods (blue dots). The results from the three methods are consistent.\label{fig:x2}}
\end{center}
\end{figure}
The quark momentum fraction in the nucleon can be calculated with the traceless part of the energy momentum tensor, and it should be consistent between calculations with two different operators.
The first one uses  the combination of the diagonal temporal and spatial components of the energy momentum tensor,
\bea
\langle x \rangle^{\textrm{E}}&\equiv&\frac{\textrm{Tr}[\Gamma^e\langle P |\int d^3x O^{\textrm{E}}(x) |P \rangle]}{E\textrm{Tr}[\Gamma^e\langle P|P \rangle]}.
\eea
where $O^{\textrm{E}}(x)=\overline{\psi}(x) \frac{1}{2}(\gamma_4\overleftrightarrow{D}_4 -\frac{1}{3}\sum_{i=1,2,3} \gamma_i\overleftrightarrow{D}_i) \psi (x)$ is the traceless part of the energy momentum tensor $T_{44}$ and is a measure of the quark fraction of the nucleon mass or
energy. The related matrix element can be calculated in the rest frame and, as a result, it will have a good signal. On the other hand, the operator $T_{44}$ itself can have mixing with lower dimension operators like the dimension-3 scalar operator $\overline{\psi}(x)\psi (x)$. Nevertheless, such a mixing will be canceled due the subtraction of the diagonal spatial components in $O_{\textrm{E}}$.

The other approach uses the forward off-diagonal matrix components of the energy momentum tensor ($T_{4i}$) in a moving frame,
\bea
\langle x \rangle^{\textrm{P}}&\equiv&\frac{\textrm{Tr}[\Gamma^e\langle P |\int d^3x \overline{\psi}(x) \frac{1}{4}(\gamma_i\overleftrightarrow{D}_4+ \gamma_4\overleftrightarrow{D}_i) \psi(x) |P \rangle]}{p_i\textrm{Tr}[\Gamma^e\langle P|P \rangle]}\nonumber\\
\eea
with $p_i$ being the $i$-th component of the nucleon momentum. Therefore, it is a measure of the quark momentum fraction in a moving nucleon. Such a scheme is free of mixing of the lower dimension operators due to its tensor structure, while the corresponding matrix element is proportional to the momentum and is thus more noisy than that from the first approach, because mixed momentum sources are involved for the matrix element of the nucleon at non-zero momentum.

Fig.~\ref{fig:x0} shows the plateau fit values of the $t_{2}$=10 case for the quark isovector momentum fraction. They are $\langle x\rangle^{\textrm{E}}$ from the diagonal components of
the energy-momentum tensor with the nucleon in the rest frame and also $\langle x\rangle^{\textrm{P}}$ 
from the off-diagonal components in a moving frame with different momenta. The results from 
both the diagonal and off-diagonal components (and also those from different momenta) are consistent, but $\langle x\rangle^{\textrm{E}}$ provides much better SNR. The sink-source separation dependence is shown in Fig.~\ref{fig:x1}, for both results based on the diagonal components and off-diagonal components. It is interesting to observe that the separation dependence of the isovector quark momentum fraction based on the off-diagonal components seems to be milder than that based on the diagonal ones, for the cases with $t_{2}$=8 and 10. The $\langle x\rangle^{\textrm{P}}$ case with $t_{2}$=12 seems to have some $t$ dependence at the smeared source end, but it could be due to the statistical fluctuation due to relatively poor signal.

As in Ref.~\cite{mike2015}, the renormalization factor for the ensemble we used has been obtained with the one-loop lattice perturbative theory,
as 1.049(3), in the $\overline{MS}$ scheme at 2 GeV. The error is from the uncertainty of the lattice spacing. The renormalized values of the isovector quark momentum fraction of $\langle x\rangle_{\textrm{E}}$ from the three fitting methods are plotted in Fig.~\ref{fig:x2}, and those at the unitary point are listed in 
Table \ref{table:axial}.

\section{Summary}\label{sec:summary}

 We have introduced a new method to calculate the nucleon matrix elements in the connected
 insertion. The stochastic sandwich method (SSM) with low-mode substitution (LMS) is an approach
 which uses low modes for the all-to-all quark propagator between the current and the sink and
 the corresponding high-mode contribution is taken care of by the noise propagator from the sink to
 the current. We have shown that it is more efficient than the sink- and current- sequential methods.
  However, it does not scale well with volume which requires more low eigenmodes. It will lose its advantage
 when the overhead from calculating the LMS for all the quark propagators involved is more than the amount it saves compared with  the sink-sequential or current-sequential method. But this will occur only at volumes much larger than that used here.
 
We have used three fitting methods. One is a two-state fitting
 including the contamination from the excited-state transition and the second is the summed-slope method. The third is a mix of these two methods.
 
The proton isovector axial-vector coupling $g_A^3$ we obtain with the overlap fermion at the unitary point with $m_{\pi}$=330 MeV is 
\begin{equation}
g_A^3 = 1.166(19) 
\end{equation}
which is is $\sim 8\%$ smaller than the experimental value.

The separation dependence of this quantity is mild. Since it is smaller than the experimental value on this
lattice, it is essential to repeat the calculation of $g_A^3$ on larger volumes and with lighter quark masses.

For the isovector scalar matrix element in the proton, the renormalized value at $\overline{MS}$(2GeV) at the unitary point is 
\begin{equation}
g_S^3 = 0.74(4). 
\end{equation}
This shows that, despite the fact that there are 2 $u$ valence quarks and only one $d$ quark in the proton, the $d$ contribution to the scalar matrix element per quark is more than that of the $u${, as 
\bea
\frac{g_{S,CI}^u}{2g_{S,CI}^d}=0.67(2)
\eea
is much smaller than one}. This has been interpreted~\cite{Liu:1993cv} to be related to the Gottfried sum rule 
violation~\cite{Amaudruz:1991at} where it is found experimentally that there are more $d$ antipartons than $u$ antipartons.

In the isovector quark momentum fraction case, the bare value we obtained at the unitary point on the ensemble mentioned above is 
\begin{equation}
\langle x\rangle_{u-d}=0.192(19),
\end{equation}
with the renormalization factor 1.049(3) from one-loop lattice perturbative theory \cite{mike2015}. This value is similar to {those from most lattice calculations} \cite{Alexandrou:2013joa,Aoki:2010xg,Bali:2014gha,Pleiter:2011gw,Syritsyn:2014xwa} and is larger than the experimental value.
However, the $O(a^2)$ error has not been considered. It can be assessed by imposing the momentum and angular momentum sum rules at finite lattice spacing as is demonstrated in a quenched calculation~\cite{Deka:2013zha}. We will return to this issue when the complete lattice simulation of the momentum and angular-momentum decompositions is carried out.

We will perform calculations with physical sea quark masses in the future.

\section*{Acknowledgments}

We thank the RBC and UKQCD Collaborations for providing us their DWF gauge configurations. This work is supported in part by the U.S. Department of Energy under Grant No.\ DE-FG05-84ER40154, and DE-SC0013065. A.A. acknowledges the support of NSF CAREER through grant PHY-1151648. M.G. is partially supported by the National Science Foundation of China (NSFC) under the project No.\ 11405178 and the Youth Innovation Promotion Association of CAS (2015013). This research used resources of the Oak Ridge Leadership Computing Facility at the Oak Ridge National Laboratory, which is supported by the Office of Science of the U.S. Department of Energy under Contract No. DE-AC05-00OR22725.

\end{document}